\documentclass[10pt,twocolumn,twoside]{IEEEtran}

\pdfoutput=1

\usepackage{amsmath,graphicx}

\usepackage{latexsym}
\usepackage{graphics}
\usepackage{amssymb}
\usepackage{cite}
\usepackage{array}
\usepackage{url}
\usepackage{subfigure}
\usepackage{color}

\def\Pr{\mathrm{Pr}}

\def\id{\tt{id}}
\def\nid{\tt{ni}}

\newcommand{\condProdtwo}[2]{\overset{#2}{\underset{#1}{\prod}}}
\newcommand{\fracSumtwo}[2]{\overset{#2}{\underset{#1}{\sum}}}

\newtheorem{remark}{Remark}
\newtheorem{theorem}{Theorem}
\newtheorem{corollary}{Corollary}
\newtheorem{lemma}{Lemma}

\newtheorem{proposition}{Proposition}

\normalsize

\begin{document}

\title{A New Look at Dual-Hop Relaying: \\
Performance Limits with Hardware Impairments}

\author{Emil~Bj\"ornson,~\IEEEmembership{Member,~IEEE,}
        Michail~Matthaiou,~\IEEEmembership{Senior~Member,~IEEE,}
        and~M\'erouane~Debbah,~\IEEEmembership{Senior~Member,~IEEE}
\thanks{\copyright 2013 IEEE. Personal use of this material is permitted. Permission from IEEE must be obtained for all other uses, in any current or future media, including reprinting/republishing this material for advertising or promotional purposes, creating new collective works, for resale or redistribution to servers or lists, or reuse of any copyrighted component of this work in other works.}%
\thanks{Manuscript received April 16, 2013; revised July 24, 2013. The editor coordinating the review of this paper and approving it for publication was J. Wang. Supplementary downloadable material is available at https://github.com/emilbjornson/new-look-at-relaying, provided by the authors. The material includes Matlab code that reproduces all simulation results.}%
\thanks{E.~Bj\"ornson and M.~Debbah are with the Alcatel-Lucent Chair on Flexible Radio, SUPELEC, Gif-sur-Yvette, France (e-mail: \{emil.bjornson,merouane.debbah\}@supelec.fr). E.~Bj\"ornson is also with the ACCESS Linnaeus Centre, Signal Processing Lab, KTH Royal Institute of Technology, Stockholm, Sweden.}%
\thanks{M.~Matthaiou is with the Department of Signals and Systems, Chalmers University of Technology, 412 96, Gothenburg, Sweden (e-mail: michail.matthaiou@chalmers.se).}
\thanks{Parts of this work were published at the IEEE Conference on Acoustics, Speech, and Signal Processing (ICASSP), Vancouver, Canada, May 2013 \cite{Bjornson2013icassp}.}%
\thanks{E.~Bj\"ornson is funded by the International Postdoc Grant 2012-228 from The Swedish Research Council. This research has been supported by the ERC Starting Grant  305123 MORE (Advanced Mathematical Tools for Complex Network Engineering). The work of M.~Matthaiou has been supported in part by the Swedish Governmental Agency for Innovation Systems (VINNOVA) within the VINN Excellence Center Chase.}%
\thanks{Digital Object Identifier 10.1109/TCOMM.2013.100913.130282}}

\maketitle

\begin{abstract}
Physical transceivers have hardware impairments that create distortions which degrade the performance of communication systems. The vast majority of technical contributions in the area of relaying neglect hardware impairments and, thus, assumes ideal hardware. Such approximations make sense in low-rate systems, but can lead to very misleading results when analyzing future high-rate systems. This paper quantifies the impact of hardware impairments on dual-hop relaying, for both amplify-and-forward and decode-and-forward protocols. The outage probability (OP) in these practical scenarios is a function of the effective end-to-end signal-to-noise-and-distortion ratio (SNDR). This paper derives new closed-form expressions for the exact and asymptotic OPs, accounting for hardware impairments at the source, relay, and destination. A similar analysis for the ergodic capacity is also pursued, resulting in new upper bounds. We assume that both hops are subject to independent but non-identically distributed Nakagami-$m$ fading. This paper validates that the performance loss is small at low rates, but otherwise can be very substantial. In particular, it is proved that for high signal-to-noise ratio (SNR), the  end-to-end SNDR converges to a deterministic constant, coined the \emph{SNDR ceiling}, which is inversely proportional to the level of impairments. This stands in contrast to the ideal hardware case in which the end-to-end SNDR grows without bound in the high-SNR regime. Finally, we provide fundamental design guidelines for selecting hardware that satisfies the requirements of a practical relaying system.
\end{abstract}


\IEEEpeerreviewmaketitle
\markboth{IEEE Transactions on Communications, VOL.~61, NO.~11, NOVEMBER 2013}{BJ\"ORNSON $\textrm{et al.}$: A NEW LOOK AT DUAL-HOP RELAYING: PERFORMANCE LIMITS WITH HARDWARE IMPAIRMENTS}%

\section{Introduction}
\IEEEPARstart{T}{he} use of relay nodes for improving coverage, reliability, and quality-of-service in wireless systems has been a hot research topic over the past decade, both in academia \cite{Laneman2004a,Hasna2004a,Hasna2004b} and in industry \cite{Yang2009,Hua2012}. This is due to the fact that, unlike macro base stations, relays are low-cost nodes that can be easily deployed and, hence, enhance the network agility. The vast majority of works in the context of relaying systems make the standard assumption of ideal transceiver hardware.

However, in practice, hardware suffers from several types of impairments; for example, phase noise, I/Q imbalance, and high power amplifier (HPA) nonlinearities among others \cite{Costa2002a,Schenk2008a,Studer2010a}. The impact of hardware impairments on various types of single-hop systems was analyzed in
\cite{Costa2002a,Schenk2008a,Studer2010a,wenk2010mimo,Zetterberg2011a,schenk2007IQ,Dardari2000a,Bjornson2012b,Samuel2008a,Riihonen2010a,Qi2012a,Bjornson2013c,Bjornson2013d}.
For instance, I/Q imbalance was considered in \cite{schenk2007IQ} and it was shown to attenuate the amplitude and rotate the phase of the desired constellation. Moreover, it creates an additional image-signal from the mirror subcarrier, which leads to a symbol error rate floor. In addition, \cite{Dardari2000a} characterized the effect of non-linear HPAs as a distortion of the constellation position plus an additive Gaussian noise. The authors therein showed that, in the presence of HPA non-linearities, the bit-error-rate increases compared to linear HPAs; for severe non-linearities, an irreducible error floor emerges. Hardware impairments are typically mitigated by compensation algorithms, but there are always residual impairments \cite{Schenk2008a,Studer2010a,wenk2010mimo}.
As a general conclusion, hardware impairments have a deleterious impact on the achievable performance \cite{Zetterberg2011a,schenk2007IQ,wenk2010mimo,Dardari2000a,Bjornson2012b,Bjornson2013c,Bjornson2013d,Samuel2008a,Riihonen2010a,Qi2012a}. This effect is more pronounced in high-rate systems, especially those employing inexpensive hardware \cite{Schenk2008a}. Recent works in information theory have demonstrated that non-ideal hardware severely affects multi-antenna systems; more specifically, \cite{Bjornson2013c} proved that there is a finite capacity limit at high {signal-to-noise ratio (SNR)}, while \cite{Bjornson2013d} provided a general resource allocation framework where existing signal processing algorithms are redesigned to account for impairments.

Despite the importance of transceiver hardware impairments, their impact on one-way relaying\footnote{Analysis of two-way AF relaying was conducted in our recent paper \cite{Matthaiou2013a}.} has only been partially investigated; bit error rate simulations were conducted in \cite{Samuel2008a} for amplify-and-forward (AF) relaying, while \cite{Riihonen2010a,Qi2012a} derived expressions for the bit/symbol error rates considering only non-linearities or I/Q imbalance, respectively. Most recently,
\cite{Naofal1,Naofal2} elaborated on the impact of I/Q imbalance on AF relaying and suggested novel digital baseband compensation algorithms.
In this paper, we follow a different line of reasoning by providing a detailed performance analysis of dual-hop relaying systems in the presence of \textit{aggregate transceiver impairments}, both for AF and decode-and-forward (DF) protocols. To the best of our knowledge, this is the first paper presenting an analytical study of relaying with transceiver impairments under the generalized system model of \cite{Schenk2008a,Studer2010a,wenk2010mimo,Zetterberg2011a}. The paper makes the following specific contributions:
\begin{itemize}
	\item We introduce a general model to account for transceiver hardware impairments in relaying. Unlike the works of \cite{Riihonen2010a,Qi2012a,Naofal1,Naofal2}, which examined the impact of a single type of impairments, we herein take a macroscopic look and investigate the aggregate impact of hardware impairments.

	\item After obtaining the instantaneous end-to-end signal-to-noise-plus-distortion ratios (SNDRs) for both AF and DF relaying,
	we derive new closed-form expressions for the exact outage probability (OP) of the system. This enables us to characterize the impact of impairments for any arbitrary SNR value. New upper bounds on the ergodic capacity are also provided. Note that our analysis considers Nakagami-$m$ fading, which has been extensively used in the performance analysis of communication systems.

	\item In order to obtain more engineering insights, we elaborate on the high-SNR regime and demonstrate the presence of a so-called SNDR ceiling. This fundamental ceiling is explicitly quantified and its value is shown to be inversely proportional to the level of impairments. This observation manifests that both AF and DF relaying systems are intimately limited by hardware impairments---especially at high SNRs and when high rates are desirable. On a similar note, the ergodic capacity exhibits a so-called capacity ceiling.

	\item In the last part of the paper, we provide some design guidelines for optimizing the performance of hardware-constrained relaying systems. These results are of particular importance when it comes down to finding the \emph{lowest} hardware quality (i.e., \emph{highest} level of impairments) that can theoretically meet stipulated requirements.
\end{itemize}

The remainder of the paper is organized as follows: In Section \ref{section-system-model}, the signal and system models, for both ideal and impaired hardware,
are outlined. For the sake of generality, we consider both dual-hop AF and DF relaying and assume that both hops are subject to independent and non-identically distributed fading. In Section \ref{section_performance_analysis}, an OP analysis is pursued that can be applied for any type of fading and is specialized to the cases of Nakagami-$m$ and Rayleigh fading. A similar analysis for the ergodic capacity is performed in Section \ref{sec:capacity}, which results in new upper bounds. The performance limits of hardware-constrained relaying systems in the high-SNR regime are examined in Section \ref{section_asymptotics} and some fundamental design guidelines are also obtained. Our numerical results are provided in Section \ref{sec:results}, while Section \ref{sec:conclusions} concludes the paper.

\subsection{Notation}
Circularly-symmetric complex Gaussian distributed variables are denoted as $x \sim \mathcal{CN}(a,b)$ where $a$ is the mean value and $b>0$ is the variance.
Gamma distributed variables are denoted as $\rho \sim \mathrm{Gamma}(\alpha,\beta)$, where $\alpha \geq 0$ is the shape parameter and $\beta>0$ is the scale parameter.
The expectation operator is denoted $\mathbb{E}\{\cdot\}$ and $\Pr \{ \mathcal{A} \}$ is the probability of an event $\mathcal{A}$. The operator $\triangleq$ denotes a definition.
The gamma function $\Gamma(n)$  of an integer $n$ satisfies $\Gamma(n) = (n-1)!$.

\section{Signal and System Model}\label{section-system-model}

\begin{figure}[tbp!]
\hfill
\subfigure[Classical dual-hop relaying with ideal transceiver]{\label{figure_block-model1} \includegraphics[width=\columnwidth]{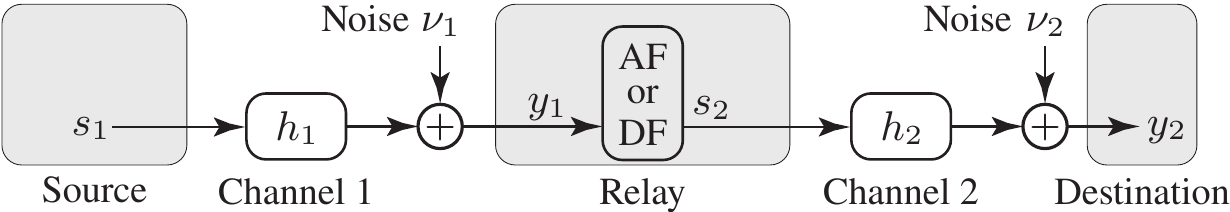}}
\hfill
\subfigure[Generalized dual-hop relaying with hardware impairments]{\label{figure_block-model2} \includegraphics[width=\columnwidth]{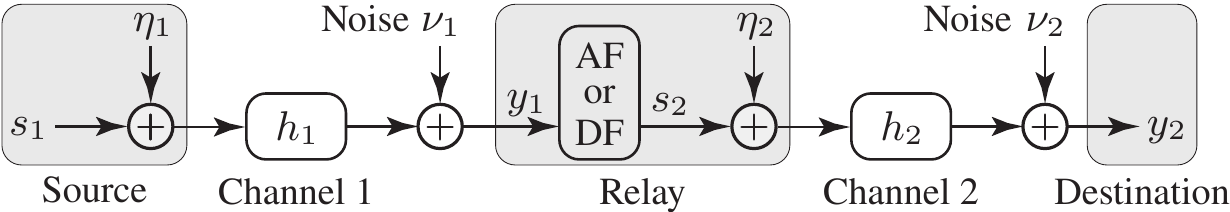}}
\hfill
\caption{Block diagram of AF/DF relaying with (a) ideal hardware or (b) non-ideal hardware with transceiver impairments modeled by the aggregate distortion noises $\eta_1,\eta_2$.}  \label{figure_block-model}
\end{figure}

This paper revisits classical dual-hop relaying where a source communicates with a destination through a relay; see Fig.~\ref{figure_block-model}(a). There is no direct link between the source and the destination (e.g., due to heavy shadowing), but the results herein can be extended to that scenario as well. Contrary to most prior works, we consider a generalized system model that accounts for transceiver hardware impairments. This model is described in the following subsections and the block model is shown in Fig.~\ref{figure_block-model}(b).

\subsection{Preliminaries on Distortion Noise from Impairments}
\label{subsec:preliminaries_distortion}

We first describe a generalized system model for single-hop transmission that originates from \cite{Schenk2008a,Studer2010a,wenk2010mimo,Zetterberg2011a}.
Suppose an information signal $s \in \mathbb{C}$ is conveyed over a flat-fading wireless channel $h \in \mathbb{C}$ with additive noise $\nu \in \mathbb{C}$.  This channel can, for example, be one of the subcarriers in a multi-carrier system based on orthogonal frequency-division multiplexing (OFDM) \cite{Zhu2009a}.
The received signal is conventionally modeled as
\begin{equation}  \label{eq_classical_model}
y = hs + \nu
\end{equation}
where $h$, $s$, and $\nu$ are statistically independent. However, physical radio-frequency (RF) transceivers suffer from impairments that are not accurately captured in this way. Informally speaking, such impairments 1) create a mismatch between the intended signal $s$ and what is actually generated and emitted; and 2) distort the received signal during the reception processing. This calls for the inclusion of additional distortion noise sources that are statistically dependent on the signal power and channel gain.

Detailed distortion models are available for different sources of impairments (e.g., I/Q imbalance, HPA non-linearities, and phase-noise);  see \cite{Schenk2008a} for a detailed description of hardware impairments in OFDM systems and related compensation algorithms. However, the combined influence at a given flat-fading subcarrier is often well-modeled by a generalized channel model \cite{Schenk2008a}, where the received signal becomes
\begin{equation}  \label{eq_generalized_model}
y = h (s+\eta_{\tt{t}}) + \eta_{\tt{r}} + \nu
\end{equation}
while $\eta_{\tt{t}},\eta_{\tt{r}}$ are \emph{distortion noises} from impairments in the transmitter and receiver, respectively \cite{Schenk2008a}. The distortion noises are defined as
\begin{equation} \label{eq_model_impairments}
 \eta_{{\tt t }} \sim \mathcal{CN}(0, \kappa_{{\tt t }}^2 P ), \quad \eta_{{\tt r}} \sim \mathcal{CN}(0, \kappa_{{\tt r }}^2 P |h|^2 )
\end{equation}
which is a model that has been supported and validated by many theoretical investigations and measurements (see e.g., \cite{Dardari2000a,Priyanto2007a,Studer2010a,wenk2010mimo,Zetterberg2011a} and references therein). The design parameters $\kappa_{{\tt t }},\kappa_{{\tt r }} \geq 0$ are described below. The joint Gaussianity in \eqref{eq_model_impairments} is explained by the aggregate effect of many impairments.\footnote{Note that the Gaussian assumption holds particularly well for the residual distortion when compensation algorithms are applied to mitigate hardware impairments \cite{Studer2010a}.} For a given channel realization $h$, the aggregate distortion seen at the receiver has power
\begin{equation} \label{eq:sum_of_impairments}
\mathbb{E}_{\eta_{\tt{t}},\eta_{\tt{r}}}\{ |h \eta_{\tt{t}} + \eta_{\tt{r}} |^2\} = P |h|^2 (\kappa_{{\tt t }}^2 + \kappa_{{\tt r }}^2).
\end{equation}
Thus, it depends on the average signal power $P = \mathbb{E}_s\{ |s|^2 \}$ and the instantaneous channel gain $|h|^2$. Note that this dependence is not supported by the classical channel model in \eqref{eq_classical_model}, because the effective distortion noise is correlated with the channel and is not Gaussian distributed.\footnote{The effective distortion noise can be seen as two independent jointly Gaussian variables $\eta_{{\tt t }}$ and $\eta_{{\tt r}} / h$ that are multiplied with the fading channel $h$. The effective distortion noise is thus only complex Gaussian distributed when conditioning on a channel realization, while the true distribution is the product of the complex Gaussian distribution of the distortion noise and the channel fading distribution. This becomes a \emph{complex double Gaussian} distribution under Rayleigh fading \cite{Donoughue2012a}, while the distribution under Nakagami-$m$ fading does not admit any known statistical characterization.}

The design parameters $\kappa_{{\tt t }},\kappa_{{\tt r }} \geq 0$ characterize the \emph{level of impairments} in the transmitter and receiver hardware, respectively. These parameters are interpreted as the error vector magnitudes (EVMs). EVM is a common quality measure of RF transceivers and is the ratio of the average distortion magnitude to the average signal magnitude.\footnote{The EVM at the transmitter is defined as $\sqrt{\mathbb{E}_{\eta_{\tt{t}}}\{ |\eta_{\tt{t}}|^2 \} / \mathbb{E}_s \{|s|^2\}}$ \cite{EVM2005}.  3GPP LTE has EVM requirements in the range $\kappa_{\tt{t}} \in [0.08,0.175]$, where smaller values are needed to support the highest spectral efficiencies \cite[Sec.~14.3.4]{Holma2011a}.} Since the EVM measures the joint impact of different hardware impairments and compensation algorithms, it can be measured directly in practice (see e.g., \cite{EVM2005}). As seen from \eqref{eq:sum_of_impairments} it is sufficient to characterize the \emph{aggregate level of impairments} $\kappa = \sqrt{\kappa_{{\tt t }}^2 + \kappa_{{\tt r }}^2}$ of the channel, without specifying the exact contribution from the transmitter hardware ($\kappa_{{\tt t }}$) and the receiver hardware ($\kappa_{{\tt r}}$). This observation is now formalized.

\begin{lemma} \label{lemma:single-parameter_characterization}
The generalized channel in \eqref{eq_generalized_model} is equivalent to
\begin{equation}  \label{eq_generalized_model_revised}
y = h (s+\eta)+ \nu
\end{equation}
where the independent distortion noise $\eta \sim \mathcal{CN}(0, \kappa^2 P )$ describes contributions from hardware impairments at both the transmitter and the receiver, such that  $\kappa \triangleq \sqrt{\kappa_{{\tt t }}^2 + \kappa_{{\tt r }}^2}$.
\end{lemma}

The single-parameter characterization in Lemma \ref{lemma:single-parameter_characterization} is used henceforth for the sake of brevity and without loss of generality.
Note that \eqref{eq_generalized_model_revised} reduces to the classical model in \eqref{eq_classical_model} when $\kappa=0$, which represents ideal transmitter and receiver hardware since it implies that $\kappa_{{\tt t }}=\kappa_{{\tt r }}=0$.

\subsection{System Model: Relaying with Non-Ideal Hardware}

Consider the dual-hop relaying scenario in Fig.~\ref{figure_block-model}. Let the transmission parameters between the source and the relay have subscript 1 and between relay and destination have subscript 2. Using the generalized system model in Lemma \ref{lemma:single-parameter_characterization}, the received signals at the relay and destination are
\begin{equation}  \label{eq_generalized_model_relaying}
y_i = h_i (s_i+\eta_i) + \nu_i, \quad \quad i=1,2
\end{equation}
where $s_1, s_2 \in \mathbb{C}$ are the transmitted signals from the source and relay, respectively, with
average signal power $P_i=\mathbb{E}_{s_i}\{|s_i|^2\}$.
In addition, $\nu_i \sim \mathcal{CN}(0,N_i)$ represents the complex Gaussian receiver noise and $\eta_i \sim \mathcal{CN} (0, \kappa_i^2{P}_i )$ is the distortion noise (introduced in Section \ref{subsec:preliminaries_distortion}), for $i=1,2$.
The distortion noise from hardware impairments (after conventional compensation algorithms have been applied) acts as an unknown noise-like interfering signal $\eta_i$ that goes through the same channel $h_i$ as the intended signal, thus making \eqref{eq_generalized_model_relaying} fundamentally different from a conventional multiple-access channel, where each user signal experiences independent channel fading.

The channel magnitudes $|h_i|$ are modeled as independent but non-identically distributed Nakagami-$m$ variates, such that the channel gains $\rho_i \triangleq |h_i|^2 \sim \mathrm{Gamma}(\alpha_i,\beta_i)$. These are Gamma distributed with integer\footnote{The assumption of integer shape parameters is made to facilitate the, otherwise tedious, algebraic manipulations for the Nakagami-$m$ fading case.} shape parameters $\alpha_i \geq 1$ and arbitrary scale parameters $\beta_i>0$.\footnote{We recall that Nakagami-$m$ fading reduces to the classical Rayleigh fading with variance $\Omega_i$ when $\alpha_i=1$ and $\beta_i = \Omega_i$; thus, Nakagami-$m$ fading brings more degrees-of-freedom for describing practical propagation environments and has been shown to provide better fit with real measurement results in various multipath channels \cite{Sheikh1993}.}
In this case, the cumulative distribution functions (cdfs) and probability distribution functions (pdfs) of the channel gains, $\rho_i$, are
\begin{align} \label{eq_cdf_exp}
F_{\rho_i}(x) &= 1 - \sum_{j=0}^{\alpha_i -1} \frac{e^{- \frac{x}{\beta_i}}}{j!} \left( \frac{x}{\beta_i} \right)^{j}, \quad x \geq 0 \\
f_{\rho_i}(x) &= \frac{x^{\alpha_i-1} e^{- \frac{x}{\beta_i}}}{\Gamma(\alpha_i) \beta_i^{\alpha_i}}, \quad x \geq 0 \label{eq_pdf_exp}
\end{align}
for $i=1,2$. Note that most of the analysis in this paper is generic and applies for any fading distribution. The choice of Nakagami-$m$ fading is only exploited for deriving closed-form expressions for quantities such as the OP and ergodic capacity. For any fading distribution, the quantity
\begin{equation} \label{eq:SNR-definition}
{\tt SNR }_i =\frac{P_i \mathbb{E}_{\rho_i}\{ \rho_i \}  }{N_i}
\end{equation}
is referred to as the \emph{average SNR}, for $i=1,2$. The average fading power is $\mathbb{E}_{\rho_i}\{ \rho_i \} = \alpha_i \beta_i$ under Nakagami-$m$ fading.

\begin{remark}[High SNR] \label{remark:high-snr}
The level of impairment $\kappa_i$ depends on the SNR \cite{Zetterberg2011a,Studer2011a,Bjornson2013d}. In most of our analysis, we consider an arbitrary fixed ${\tt SNR }_i$ and thus $\kappa_i$ can be taken as a constant. However, some remarks are in order for our high-SNR analysis in Section \ref{section_asymptotics}. As seen from \eqref{eq:SNR-definition}, a high SNR can be achieved by having high signal power $P_i$ and/or high fading power $\mathbb{E}_{\rho_i}\{ \rho_i \} $.
If we increase the signal power to operate outside the dynamic range of the power amplifier, then the level of impairments $\kappa_i$ increases as well due to the HPA nonlinearities \cite{Studer2011a}. Advanced dynamic power adaptation is then required to maximize the performance \cite{Bjornson2012b}. If we, on the other hand, increase the fading power (e.g., by decreasing the propagation loss) then it has no impact on $\kappa_i$. For brevity, we keep the analysis clean by assuming that any change in SNR is achieved by a change in the average fading power, while the signal power is fixed.  We stress that the upper bounds and necessary conditions derived in Section \ref{section_asymptotics} are also valid when the signal power is increased, but then they will be optimistic and no longer tight in the high-SNR regime.
\end{remark}

In the next subsections, we derive the end-to-end SNDRs for AF and DF relaying, respectively.

\subsection{End-to-End SNDR: Amplify-and-Forward Relaying}

The information signal $s_1$ should be acquired at the destination. In the AF relaying protocol, the transmitted signal $s_2$ at the relay is simply an amplified version of the signal $y_1$ received at the relay: $s_2 = G\,y_1$ for some amplification factor $G>0$. With non-ideal ({\scriptsize ${\tt  ni}$})
hardware, as described by \eqref{eq_generalized_model_relaying}, the received signal at the destination is now obtained as
 \begin{align} \label{eq_channel_model_impairments}
y_2 & = h_2\, G_{\nid} \Big( h_1\, (s_1+ \eta_1) + \nu_1 \Big)+h_2\,\eta_2 + \nu_2 \\ \nonumber
& = G_{\nid}\,h_1\,h_2\,s_1 + G_{\nid}\,h_1\,h_2\,\eta_1 + G_{\nid}\,h_2\,\nu_1 + h_2\,\eta_2 + \nu_2
\end{align}
where the amplification factor $G_{\nid}$ is selected at the relay to satisfy its power constraint. The source needs no channel knowledge. If the relay has instantaneous knowledge of the fading channel, $h_1$, it can apply variable gain relaying with $G^{\tt{v}}\triangleq \sqrt{{P_2}/{\mathbb{E}_{s_1,\nu_1,\eta_1}\{ |y_1|^2\}}}$ \cite{Emamian2002a}. Otherwise, fixed gain relaying with $G^{\tt{f}}\triangleq \sqrt{{P_2}/{\mathbb{E}_{s_1,\nu_1,\eta_1,h_1}\{ |y_1|^2\}}}$ can be applied using only statistical channel information \cite{Hasna2004a}.\footnote{The relay then has a long-term power constraint $P_2=\mathbb{E}\{|G^{\tt{f}} y_1|^2\}$ where expectation is taken over signal, noise, and channel fading realizations.} For fixed and variable gain relaying, $G_{\nid}$ reads respectively as
\begin{align}
G_{\nid}^{\tt{f}}  &\triangleq \sqrt{ \frac{P_2}{P_1\, \mathbb{E}_{\rho_1}\{ \rho_1 \} (1+\kappa_1^2)+ N_1}} \\
G_{\nid}^{\tt{v}}  &\triangleq \sqrt{ \frac{P_2}{P_1\, \rho_1 (1+\kappa_1^2)+ N_1}}
\end{align}
where $\mathbb{E}_{\rho_1}\{ \rho_1 \} = \alpha_1 \beta_1$ for Nakagami-$m$ fading.

Note that variable gain relaying has always  an output power of $P_2$ at the relay, whilst for fixed gain relaying this
varies with the channel gain of the first hop. This, in turn, affects the variance of the distortion noise $\eta_2$ for the second hop, which by definition is $\mathbb{E}\{|\eta_2|^2\} = \kappa_2^2\, G_{\nid}^2 \mathbb{E}_{s_1,\nu_1}\{|y_1|^2\}$ for AF relaying. This reduces to the simple expression $\kappa_2^2\, P_2$ for variable gain relaying, while it becomes $\big(G_{\nid}^{\tt{f}}\big)^2 \! \kappa_2^2\,(P_1\,\rho_1 (1+\kappa_1^2) + N_1)$ for fixed gain relaying.

After some algebraic manipulations (e.g., using the expressions for $G_{\nid}^{\tt{v}}$), the end-to-end SNDRs for fixed and variable gain relaying are obtained as
\begin{align} \label{eq_SNDR_AF_impairments_fixed}
\gamma_{\nid}^{\tt{AF} \textrm{-} \tt{f}} & = \frac{\,\rho_1\,\rho_2}{\,\rho_1\,\rho_2\,d + \rho_2 (1+\kappa_2^2) \frac{N_1}{P_1}  + \frac{N_2}{P_1 \left(G_{\nid}^{\tt{f}}\right)^2}}\\
\gamma_{\nid}^{\tt{AF}  \textrm{-} \tt{v}} & = \frac{\rho_1\,\rho_2}{\rho_1\,\rho_2\,d \!+\! \rho_1 (1\!+\!\kappa_1^2) \frac{N_2}{ P_2} \!+\! \rho_2 (1\!+\!\kappa_2^2) \frac{N_1}{P_1} \!+\! \frac{N_1 N_2}{P_1 P_2}} \label{eq_SNDR_AF_impairments_variable}
\end{align}
respectively, assuming that the destination knows the two channels and the statistics of the receiver and distortion noises. Note that the parameter $d \triangleq \kappa_1^2+\kappa_2^2+ \kappa_1^2 \,\kappa_2^2$ that appears in \eqref{eq_SNDR_AF_impairments_fixed}--\eqref{eq_SNDR_AF_impairments_variable} plays a key  role in this paper.

\begin{remark}[Ideal Hardware]
The end-to-end SNRs for AF relaying with ideal ({\scriptsize ${\tt  id}$}) hardware were derived in \cite{Emamian2002a,Hasna2004a}. The results of this section reduce to that special case when setting $\kappa_1=\kappa_2=0$. The amplification factors then become
\begin{align}
 G_{\id}^{\tt{f}}= \sqrt{ \frac{P_2}{P_1\, \mathbb{E}_{\rho_1}\{ \rho_1 \} + N_1}}, \quad G_{\id}^{\tt{v}}= \sqrt{ \frac{P_2}{P_1\, \rho_1 + N_1}}
	\label{eq:idealgain}
\end{align}
and the end-to-end SNRs become
 \begin{equation} \label{eq_SNDR_AF_ideal}
\gamma^{\tt{AF} \textrm{-} \tt{f}}_{\id} = \frac{  \rho_1\, \rho_2}{\rho_2\, \frac{N_1}{P_1} \! +\! \frac{N_2}{P_1 (G^{\tt{f}}_{\id})^2}}, \,\,\, \gamma^{\tt{AF} \textrm{-} \tt{v}}_{\id} = \frac{ \rho_1\, \rho_2}{   \rho_1 \frac{N_2}{P_2} \!+\! \rho_2\,\frac{ N_1}{P_1} \!+\! \frac{N_1 N_2}{P_1 P_2}}
\end{equation}
for fixed and variable gain relaying, respectively. Comparing the SNDRs in \eqref{eq_SNDR_AF_impairments_fixed}--\eqref{eq_SNDR_AF_impairments_variable} with the ideal hardware case in \eqref{eq_SNDR_AF_ideal},
the mathematical form of the former is more complicated, since the product $\rho_1 \rho_2$ appears in the denominator. It is, therefore, non-trivial to generalize prior works on AF relaying with Nakagami-$m$ fading (e.g., \cite{Karagiannidis2006a,Hasna2004b,Senaratne2010a}) to the general case of non-ideal hardware. This generalization is done in Section \ref{section_performance_analysis} and is a main contribution of this paper.
\end{remark}

\subsection{End-to-End SNDR: Decode-and-Forward Relaying}

In the DF relaying protocol, the transmitted signal $s_2$ at the relay should equal the original intended signal $s_1$. This is only possible if the relay is able to decode the signal (otherwise the relayed signal is useless); thus, the effective SNDR is the minimum of the SNDRs between 1) the source and relay; and 2) the relay and destination. We assume that the relay knows $h_1$ and the destination knows $h_2$, along with the statistics of the receiver and distortion noises.

With non-ideal hardware as described by \eqref{eq_generalized_model_relaying}, the effective end-to-end SNDR becomes
\begin{equation} \label{eq_SNDR_DF_impairments}
\gamma_{\nid}^{\tt{DF}} = \min \left( \frac{P_1 \rho_1}{P_1 \rho_1 \kappa_1^2 + N_1}, \frac{P_2 \rho_2}{P_2 \rho_2 \kappa_2^2 + N_2} \right)
\end{equation}
and does not require any channel knowledge at the source. In the special case of ideal hardware (i.e., $\kappa_1=\kappa_2=0$), \eqref{eq_SNDR_DF_impairments} reduces to the classical result from \cite{Laneman2004a}; that is
\begin{equation} \label{eq_SNDR_DF_ideal}
\gamma_{\id}^{\tt{DF}} = \min \left( \frac{P_1 \rho_1}{N_1}, \frac{P_2 \rho_2}{N_2} \right).
\end{equation}
Just as for AF relaying, the SNDR expression with DF relaying is more complicated in the general case with hardware impairments.
This is manifested in \eqref{eq_SNDR_DF_impairments} by the statistical dependence between numerators and denominators, which is different from the ideal case in \eqref{eq_SNDR_DF_ideal}.

\section{Outage Probability Analysis}\label{section_performance_analysis}

This section derives new closed-form expressions for the exact OPs under the presence of transceiver impairments. These results generalize the well known results in the literature, such as \cite{Laneman2004a,Hasna2004a,Hasna2004b,Karagiannidis2006a,Senaratne2010a}, which rely on the assumption of ideal hardware.
The OP is denoted by $P_{\tt{out}}(x)$ and is the probability that the channel fading makes the effective end-to-end SNDR fall below a certain threshold, $x$, of acceptable communication quality. Mathematically speaking, this means that
\begin{equation} \label{eq_outage_prob_def}
P_{\tt{out}}(x) \triangleq \Pr \{ \gamma \leq x\}
\end{equation}
where $\gamma$ is the effective end-to-end SNDR.

\subsection{Arbitrary Channel Fading Distributions}

This subsection derives general expressions for the OP that hold true for any distributions of the channel gains $\rho_1,\rho_2$. These offer useful tools, which later will allow us to derive
closed-form expressions for the cases of Nakagami-$m$ and Rayleigh fading. Note that $\rho_1,\rho_2$ appear in both numerators and denominators of the SNDRs in \eqref{eq_SNDR_AF_impairments_fixed}--\eqref{eq_SNDR_AF_impairments_variable} and \eqref{eq_SNDR_DF_impairments}. The following lemma enable us to characterize this structure.

\begin{lemma} \label{lemma_outage_impairments}
Let $c_1,c_2,c_3$ be strictly positive constants and let $\rho$ be a non-negative random variable with cdf $F_{\rho}(\cdot)$. Then,
\begin{equation} \label{eq_lemma_probability}
\Pr \left\{ \frac{c_1 \rho}{c_2 \rho + c_3} \leq x \right\} = \begin{cases} F_{\rho} \big(\frac{c_3 x}{c_1 - c_2 x}\big), & 0 \leq x < \frac{c_1}{c_2}, \\ 1, & x \geq \frac{c_1}{c_2}. \end{cases}
\end{equation}
Suppose $c_2=0$ instead, then \eqref{eq_lemma_probability} simplifies to
\begin{equation}
\Pr \left\{ \frac{c_1 \rho}{c_3} \leq x \right\} = F_{\rho} \left( \frac{c_3 x}{c_1} \right).
\end{equation}
\end{lemma}
\begin{IEEEproof}
The left-hand side of \eqref{eq_lemma_probability} is equal to
\begin{equation}
\begin{split}
\Pr \Big\{ c_1 \rho \leq (c_2 \rho + c_3) x \Big\}
&= \Pr \left\{  \rho \leq \frac{c_3 x}{(c_1 - c_2 x)} \right\}
\end{split}
\end{equation}
after some basic algebra. The last expression is exactly $F_{\rho}\left(\frac{c_3 x}{c_1 - c_2 x}\right)$. If $(c_1 - c_2 x)\leq 0$, then the inequality is satisfied for any realization of the non-negative variable $\rho$.
\end{IEEEproof}

Based on Lemma \ref{lemma_outage_impairments}, we can derive integral expressions for the OPs with AF relaying.

\begin{proposition} \label{proposition_AF_outage_general_distribution}
Suppose $\rho_i$ is an independent non-negative random variable with cdf $F_{\rho_i}(\cdot)$ and pdf $f_{\rho_i}(\cdot)$ for $i=1,2$.
The OP with AF relaying and non-ideal hardware is
\begin{equation} \label{eq_AF_impair_outage}
\begin{split}
&P^{\tt{AF},\nid}_{\tt{out}}(x) =  \\ & 1- \int_{0}^{\infty} \!\! \bigg( 1 \!-\! F_{\rho_1} \! \bigg( \! \frac{b_2 x}{(1 \!-\! dx)} \!+\! \frac{ \frac{b_1 b_2 x^2}{1\!-\!dx} \!+ \! cx }{z (1\!-\!dx)} \bigg) \!\! \bigg) f_{\rho_2} \! \bigg( \! z \!+\! \frac{b_1 x}{1\!-\!dx} \! \bigg) d z
\end{split}
\end{equation}
for $x < \frac{1}{d}$ and $P^{\tt{AF},\nid}_{\tt{out}}(x)=1$ for $x \geq \frac{1}{d}$. Recall that $d \triangleq \kappa_1^2+\kappa_2^2+ \kappa_1^2 \,\kappa_2^2$. The choice of AF protocol determines
$b_1,b_2,c$:
\begin{equation*} \begin{cases} b_1 = 0, b_2 = \frac{N_1 (1+\kappa_2^2)}{ P_1}, c = \frac{N_2}{P_1 (G^{\tt{f}}_{\nid})^2} & \text{if fixed gain},   \\
b_1 = \frac{N_2 (1+\kappa_1^2)}{ P_2}, b_2 = \frac{N_1 (1+\kappa_2^2)}{ P_1}, c = \frac{N_1 N_2}{P_1 P_2} & \text{if variable gain}. \end{cases}
\end{equation*}
In the special case of ideal hardware, \eqref{eq_AF_impair_outage} reduces to
\begin{equation} \label{eq_AF_impair_outage_ideal}
\begin{split}
&P^{\tt{AF},\id}_{\tt{out}}(x) \\ &= 1- \int_{0}^{\infty} \!\! \bigg( 1 \!-\! F_{\rho_1} \! \bigg( \! b_2 x \!+\! \frac{ b_1 b_2 x^2 \!+ \! cx }{z} \bigg) \!\! \bigg) f_{\rho_2} \! \bigg( \! z \!+\! b_1 x \! \bigg) d z
\end{split}
\end{equation}
where the parameters $b_1,b_2,c,d$ depend on the AF protocol:
\begin{equation*} \begin{cases} b_1 = 0, b_2 = \frac{N_1}{ P_1}, c = \frac{N_2}{P_1 (G^{\tt{f}}_{\id})^2}, d=0 & \text{if fixed gain},   \\
b_1 = \frac{N_2}{ P_2}, b_2 = \frac{N_1}{ P_1}, c = \frac{N_1 N_2}{P_1 P_2}, d=0 & \text{if variable gain}. \end{cases}
\end{equation*}
\end{proposition}
\begin{IEEEproof}
The proof follows from Lemma \ref{lemma_outage_impairments} and Lemma \ref{lemma:main-result} in Appendix \ref{appendix:lemmas}, by noting that the end-to-end SNDRs for non-ideal hardware in \eqref{eq_SNDR_AF_impairments_fixed}--\eqref{eq_SNDR_AF_impairments_variable} and ideal hardware in \eqref{eq_SNDR_AF_ideal}, are of the form in
\eqref{eq:Lambda-express} for different values of $a,b_1,b_2,c,d$.
\end{IEEEproof}

The result in Lemma \ref{lemma_outage_impairments} also allows explicit expressions for the OPs with DF relaying.

\begin{proposition} \label{proposition_DF_outage_general_distribution}
Suppose $\rho_i$ is an independent non-negative random variable with cdf $F_{\rho_i}(\cdot)$ for $i=1,2$.
The OP with DF relaying and non-ideal hardware is
\begin{equation} \label{eq_DF_impair_outage_general}
P^{\tt{DF},\nid}_{\tt{out}}(x)  = \begin{cases} 1 \!-\! \condProdtwo{i=1}{2} \bigg( \! 1-F_{\rho_i}\bigg(\frac{N_i x}{P_i (1-\kappa_i^2 x)}\bigg) \bigg), & x < \frac{1}{\delta}, \\
1, & x \geq \frac{1}{\delta}, \end{cases}
\end{equation}
with $\delta \triangleq \max(\kappa_1^2,\kappa_2^2)$. In the special case of ideal hardware, \eqref{eq_DF_impair_outage_general} reduces to
\begin{equation} \label{eq_DF_impair_outage_ideal}
P^{\tt{DF},\id}_{\tt{out}}(x) = 1 -  \prod_{i=1}^2 \bigg( 1-F_{\rho_i}\bigg(\frac{N_i x}{P_i}\bigg) \bigg).
\end{equation}
\end{proposition}
\begin{IEEEproof}
For a set of independent random variables $\xi_i$ with marginal cdfs $F_{\xi_i}(x)$, the random variable $\min_i (\xi_i)$ has cdf $1-\prod_i (1-F_{\xi_i}(x))$. The proof follows by combining this standard property with Lemma \ref{lemma_outage_impairments} and \eqref{eq_SNDR_DF_impairments}--\eqref{eq_SNDR_DF_ideal}.
\end{IEEEproof}

Note that the OP expressions in Propositions \ref{proposition_AF_outage_general_distribution} and \ref{proposition_DF_outage_general_distribution} allow the straightforward computation of the OP for any channel fading distribution, either directly (for DF) or by a simple numerical integration (for AF). In Section \ref{subsection:nakagami-fading-case}, we particularize these expressions to the cases of Nakagami-$m$ and Rayleigh fading to obtain closed-form results.

Interestingly, Propositions \ref{proposition_AF_outage_general_distribution} and \ref{proposition_DF_outage_general_distribution} show that the OP, $P_{\tt{out}}(x)$, is always 1 for $x \geq \frac{1}{d}$ when using AF and 1 for $x \geq \frac{1}{\delta}$ when using DF. Note that these results hold for any channel fading distribution and SNR; hence, there are certain SNDR thresholds that can never be crossed. This has an intuitive explanation since the SNDRs derived in Section \ref{section-system-model} are upper bounded as $\gamma_{\nid}^{\tt{AF}} \leq \frac{1}{d}$ and $\gamma_{\nid}^{\tt{DF}} \leq \frac{1}{\delta}$. We elaborate further on this fundamental property in Section \ref{section_asymptotics}.

\subsection{Nakagami-$m$ and Rayleigh Fading Channels}
\label{subsection:nakagami-fading-case}

Under ideal hardware, the OPs with fixed and variable gain AF relaying were obtained in \cite[Eq.~(9)]{Hasna2004a} and \cite[Eq.~(14)]{Emamian2002a}, respectively. These prior works considered Rayleigh fading, while closed-form expression for the case of Nakagami-$m$ fading were obtained in \cite{Karagiannidis2006a,Hasna2004b,Senaratne2010a} under ideal hardware. Unfortunately, the OP in the general AF relaying case with non-ideal hardware cannot be deduced from these prior results; for example, the general analysis in \cite{Senaratne2010a} does not handle cases when $\rho_1 \rho_2$ appears in the denominator of the SNDR expression, which is the case in \eqref{eq_SNDR_AF_impairments_fixed}--\eqref{eq_SNDR_AF_impairments_variable}.

The following key theorem provides new and tractable closed-form OP expressions in the presence of transceiver hardware impairments.

\begin{theorem} \label{theorem_AF_Nakagami_Rayleigh}
Suppose $\rho_1,\rho_2$ are independent and $\rho_i \sim \mathrm{Gamma}(\alpha_i,\beta_i)$ where $\alpha_i \geq 1$ is an integer and $\beta_i>0$ for $i=1,2$.
The OP with AF relaying and non-ideal hardware is
\begin{equation} \label{eq:general_cdf_expression_nakagami}
\begin{split}
P^{\tt{AF},\nid}_{\tt{out}} & (x)  = 1 \!-\! 2 e^{- \frac{x}{1-dx} \left(\frac{b_1}{\beta_2}+\frac{b_2}{\beta_1} \right)} \sum_{j=0}^{\alpha_1-1} \sum_{n=0}^{\alpha_2-1} \sum_{k=0}^{j} C(j,n,k) \\
&  \times \left( \frac{x}{1-dx} \right)^{\alpha_2+j} \left( b_1 b_2 + \frac{c(1-dx)}{x} \right)^{\frac{n+k+1}{2}} \\
& \times K_{n-k+1}\left( 2 \sqrt{ \frac{b_1 b_2 x^2}{\beta_1 \beta_2 (1-dx)^2} + \frac{cx}{\beta_1 \beta_2 (1-dx)} } \right)
\end{split}
\end{equation}
for $x < \frac{1}{d}$ and $P^{\tt{AF},\nid}_{\tt{out}}(x)=1$ for $x \geq \frac{1}{d}$. The $\nu$th-order modified Bessel function of the second kind
is denoted by $K_{\nu}(\cdot)$, while
\begin{equation}
C(j,n,k) \triangleq \frac{b_1^{\alpha_2-n-1} b_2^{j-k} \beta_1^{\frac{k-n-1-2j}{2}} \beta_2^{\frac{n-k+1-2\alpha_2}{2}} }{k! \, (j-k)! \,n! \,(\alpha_2-n-1)!}.
\end{equation}
The parameters $a,b_1,b_2$ depend on the choice of the AF protocol and are given in Proposition \ref{proposition_AF_outage_general_distribution}, while $d \triangleq \kappa_1^2+\kappa_2^2+ \kappa_1^2 \,\kappa_2^2$.
In the special case of Rayleigh fading ($\alpha_i=1$, $\beta_i = \Omega_i$), the OP becomes
\begin{align}\label{eq:general_cdf_expression_rayleigh}
P^{\tt{AF},\nid}_{\tt{out}} &(x)  = 1 - \frac{2 e^{- \frac{x}{1-dx} \left(\frac{b_1}{\Omega_2}+\frac{b_2}{\Omega_1} \right)}}{\sqrt{\Omega_1 \Omega_2}}  \sqrt{  \frac{ b_1 b_2 x^2}{ (1-dx)^2} + \frac{c x}{ (1-dx)}} \notag \\
& \times K_{1}\left( \frac{2}{\sqrt{\Omega_1 \Omega_2}} \sqrt{ \frac{b_1 b_2 x^2}{ (1-dx)^2} + \frac{cx}{(1-dx)} } \right)
\end{align}
for $x < \frac{1}{d}$ and $P^{\tt{AF},\nid}_{\tt{out}}(x)=1$ for $x \geq \frac{1}{d}$.
\end{theorem}
\begin{IEEEproof}
This results follows by combining Proposition \ref{proposition_AF_outage_general_distribution} with Lemma \ref{lemma:main-result} in Appendix \ref{appendix:lemmas}.
\end{IEEEproof}

Theorem \ref{theorem_AF_Nakagami_Rayleigh} generalizes the prior works mentioned above, which all assume ideal hardware. Note that OP expressions equivalent to those in prior works, can be obtained by setting $\kappa_1=\kappa_2=0$ in Theorem \ref{theorem_AF_Nakagami_Rayleigh}, which effectively removes the possibility of $x \geq \frac{1}{d}$ since $\frac{1}{d} = \infty$.

Next, closed-form OP expressions for DF relaying are obtained in the general case of non-ideal hardware.

\begin{theorem} \label{theorem_DF_Nakagami_Rayleigh}
Suppose $\rho_1,\rho_2$ are independent and $\rho_i \sim \mathrm{Gamma}(\alpha_i,\beta_i)$ where $\alpha_i \geq 1$ is an integer and $\beta_i>0$ for $i=1,2$.
The OP with DF relaying and non-ideal hardware is
\begin{equation} \label{eq_DF_impair_outage_nakagami}
P^{\tt{DF},\nid}_{\tt{out}}(x)  = 1 \!-\! \condProdtwo{i=1}{2} \Bigg( \fracSumtwo{j=0}{\alpha_i -1} \frac{e^{- \frac{N_i x}{P_i \beta_i (1-\kappa_i^2 x)}}}{j!} \left( \!\frac{N_i x}{P_i \beta_i(1-\kappa_i^2 x)} \! \right)^{j} \Bigg)
\end{equation}
for $x < \frac{1}{\delta}$ where $\delta \triangleq \max(\kappa_1^2,\kappa_2^2)$ and $P^{\tt{DF},\nid}_{\tt{out}}(x)=1$ for $x \geq \frac{1}{\delta}$.
In the special case of Rayleigh fading ($\alpha_i=1$, $\beta_i = \Omega_i$), the OP becomes
\begin{equation} \label{eq_DF_impair_outage_rayleigh}
P^{\tt{DF},\nid}_{\tt{out}}(x) = \begin{cases}
1 - e^{- \fracSumtwo{i=1}{2} \frac{N_i x}{P_i \Omega_i (1-\kappa_i^2 x)} }, & 0 \leq x < \frac{1}{d},\\
1, & x \geq \frac{1}{d}.
\end{cases}
\end{equation}
\end{theorem}
\begin{IEEEproof}
By plugging the respective cdfs of Nakagami-$m$ and Rayleigh fading into Proposition \ref{proposition_DF_outage_general_distribution}, we obtain the desired results.
\end{IEEEproof}

We stress that Theorem \ref{theorem_DF_Nakagami_Rayleigh} generalizes the classical results of \cite[Eq.~(21)]{Hasna2002} and \cite{Laneman2004a, Zhao2005}, which were reported for the case of DF relaying with ideal hardware. We also note that Theorem \ref{theorem_DF_Nakagami_Rayleigh} can be straightforwardly extended to multi-hop relaying scenarios with $M>2$ hops. The only difference would be to let the index $i \in \{1,\ldots,M\}$ account for all $M$ hops.

\vspace{-3mm}

\section{Ergodic Capacity Analysis}\label{sec:capacity}

In the case of ergodic channels, the ultimate performance measure is the ergodic channel capacity, expressed in bits/channel use. Similar to \cite{Farhadi2008a,Waqar2011,Zhong2011}, the term \emph{channel} refers to the end-to-end channel with a fixed relaying protocol (e.g., AF or DF). When compared to the ergodic capacity with arbitrary relaying protocols, as in \cite{Host2005a}, the results for the AF and DF relaying channels should be interpreted as ergodic achievable rates. This section provides tractable bounds and approximations for the ergodic capacities of AF and DF relaying.

\vspace{-3mm}

\subsection{Capacity of AF Relaying}

While the capacity of the AF relaying channel with ideal hardware has been well investigated in prior works (see e.g., \cite{Farhadi2008a,Waqar2011,Zhong2011} and references therein), the case of AF relaying with hardware impairments has been scarcely addressed. In the latter case, the channel capacity can be expressed as
\begin{equation}\label{capform}
\begin{split}
	C_{\nid}^{\tt{AF}} &\triangleq \frac{1}{2} \mathbb{E} \left\{ \log_2\left(1+\gamma_{\nid}^{\tt{AF}}\right) \right\}
\end{split}
\end{equation}
where the factor $1/2$ accounts for the fact that the entire communication occupies two time slots. The ergodic capacity can be computed by numerical integration, using the fact that the pdf of $\gamma_{\nid}^{\tt{AF}}$ can be deduced by differentiating the cdf in Theorem \ref{theorem_AF_Nakagami_Rayleigh}. However, an exact evaluation of \eqref{capform} is tedious, if not impossible, to obtain in closed-form.

To characterize the ergodic capacity of the AF relaying channel with fixed or variable gain, an upper bound is derived by the following theorem.

\begin{theorem} \label{theorem:upper-bound-capacity-AF}
For Nakagami-$m$ fading channels, the ergodic capacity $C_{\nid}^{\tt{AF}}$ (in bits/channel use) with AF relaying and non-ideal hardware is upper bounded as
\begin{align}\label{eq:cap_fixed_strict}
	C_{\nid}^{\tt{AF}} \leq \frac{1}{2} \log_2 \left( 1 + \frac{{\cal J}}{ {\cal J} d  + 1} \right)
\end{align}
with
\begin{equation} \label{eq:G-expectation}
\begin{split}
&{\cal J}
\triangleq\sum_{n=0}^{\alpha_1-1} \sum_{k=0}^{\alpha_2-1} \sum_{m=0}^{k} \sum_{q=0}^{n+m+2} \frac{ (n+1) \beta_1^{ \frac{n-m+2-2\alpha_1}{2} } \beta_2^{\frac{m-n-2k}{2}} }{ (k-m)! (\alpha_1-n-1)!  } \\
&\times \frac{ \left(\frac{b_1}{b_2}\right)^{\frac{n-m+2+2k}{2}}  \left(\frac{c}{b_1}\right)^q  }{c (-1)^{\alpha_1+k-q+1}} {n+m+2 \choose q}   \frac{d^{\alpha_1+k-q+1}}{dt^{\alpha_1+k-q+1}} \Bigg\{
e^{ \frac{c t}{2 b_1} }\\
& \times  W_{-\frac{n+m+2}{2} , \frac{n-m+1}{2} } \!\! \left( \frac{c}{2b_1} \! \left(t - \sqrt{t^2 \!-\! \frac{4b_1}{b_2 \beta_1 \beta_2}} \right) \!\!\right) \\ &\times
W_{-\frac{n+m+2}{2} , \frac{n-m+1}{2} } \!\!  \left( \frac{c}{2b_1} \! \left(t + \sqrt{t^2 \!-\! \frac{4b_1}{b_2 \beta_1 \beta_2}} \right) \!\!\right) \!\!
\Bigg\} \! \Bigg|_{t = \frac{1}{\beta_1} + \frac{b_1}{b_2 \beta_2}}
\end{split}
\end{equation}
where $W_{\cdot,\cdot}(\cdot)$ denotes the Whittaker $W$ function \cite[Ch.~9.22]{Gradshteyn2007a}. The parameters $b_1,b_2,c$ take different values for fixed and variable gain relaying and are given in Proposition~\ref{proposition_AF_outage_general_distribution}.
\end{theorem}
\begin{IEEEproof}
For brevity, the proof is given in Appendix \ref{appendix:proof-of-theorems}.
\end{IEEEproof}

Although the expression in \eqref{eq:G-expectation} is complicated, we note that analytical expressions for the derivatives of arbitrary order are known for the Whittaker $W$ function \cite{Senaratne2010a}; thus, the upper bound in Theorem \ref{theorem:upper-bound-capacity-AF} can be analytically evaluated in an efficient way. For the purpose of numerical illustrations in Section \ref{sec:results}, we implemented the upper bound in Theorem \ref{theorem:upper-bound-capacity-AF} using the Symbolic Math Toolbox in MATLAB \cite{Weiss2010a}.

Nevertheless, a simpler closed-form expression for the ergodic capacity is achieved by applying the approximation
\begin{align}\label{eq:approx_log}
\mathbb{E}\left\{\log_2\left(1+\frac{x}{y}\right)\right\} \approx  \log_2\left(1+\frac{\mathbb{E}\{x\}}{\mathbb{E}\{y\}}\right)
\end{align}
to \eqref{capform}. For Nakagami-$m$ fading channels, we obtain
\begin{align}\label{eq:cap_fixed}
	C_{\nid}^{\tt{AF}}\approx \!\frac{1}{2}\log_2\left(1\!+\!\frac{\alpha_1\alpha_2 \beta_1\beta_2}{\alpha_1\alpha_2 \beta_1\beta_2d+\alpha_1 \beta_1 b_1+\alpha_2 \beta_2 b_2  + c}\right)
\end{align}
where the parameters $b_1,b_2,c$ were defined in Proposition \ref{proposition_AF_outage_general_distribution} for fixed and variable gain relaying.
Despite the approximative nature of this result, we show numerically in Section \ref{sec:results} that \eqref{eq:cap_fixed} is an upper bound that is almost as tight as the one in Theorem \ref{theorem:upper-bound-capacity-AF}. In addition, both expressions are asymptotically exact in the high-SNR regime.

\subsection{Capacity of DF Relaying}

Next, we consider the ergodic capacity of the DF relaying channel which is more complicated to analyze than the AF relaying channel; the decoding and re-encoding at the relay gives additional constraints and degrees-of-freedom to take into account \cite{Host2005a}. For example, an information symbol must be correctly decoded at the relay before re-encoding, and different symbol lengths and transmit powers can then be allocated to the two hops to account for asymmetric fading/hardware conditions.

For brevity, we consider a strict DF protocol with fixed power and equal time allocation.
Based on \cite[Eq.~(45)]{Host2005a}, \cite[Eq.~(11a)]{Farhadi2008a}, and the effective SNDR expression in \eqref{eq_SNDR_DF_impairments}, the ergodic channel capacity under hardware impairments can be upper bounded as
\begin{equation}\label{capform_DF}
\begin{split}
	C_{\nid}^{\tt{DF}} &\leq  \min_{i=1,2} \, \frac{1}{2} \mathbb{E} \left\{ \log_2\left(1+ \frac{P_i \rho_i}{P_i \rho_i \kappa_i^2 + N_i} \right) \right\}.
\end{split}
\end{equation}
The intuition behind this expression is that the information that can be sent from the source to the destination is upper bounded by the minimum of the capacities of the individual channels.
A closed-form upper bound, which holds for any channel fading distributions, is derived in the new theorem.

\begin{theorem} \label{theorem:upper-bound-capacity-DF}
The ergodic capacity $C_{\nid}^{\tt{DF}}$ (in bits/channel use) with DF relaying and non-ideal hardware  is upper bounded as
\begin{align}\label{eq:cap_fixed_strict_DF}
	C_{\nid}^{\tt{DF}} \leq  \min_{i=1,2} \, \frac{1}{2} \log_2\left(1+ \frac{{\tt SNR }_i}{{\tt SNR }_i \kappa_i^2 + 1} \right).
\end{align}
\end{theorem}
\begin{IEEEproof}
For brevity, the proof is given in Appendix \ref{appendix:proof-of-theorems}.
\end{IEEEproof}

This theorem shows clearly the impact of hardware impairments on the channel capacity: the distortion noise shows up as an interference term that is proportional to the SNR. The upper bound will therefore not grow unboundedly with the SNR, as would be the case for ideal hardware \cite{Host2005a,Farhadi2008a}. The next section elaborates further on the high-SNR regime.

\section{Fundamental Limits: Asymptotic SNR Analysis}
\label{section_asymptotics}

To obtain some insights on the fundamental impact of impairments, we now elaborate on the high-SNR regime. Recall the SNR definition, ${\tt SNR }_i =\frac{P_i \mathbb{E}_{\rho_i}\{ \rho_i \}  }{N_i}$ for $i=1,2$, in \eqref{eq:SNR-definition} and the corresponding Remark \ref{remark:high-snr} on the SNR scaling.

For the ease of presentation, we assume that ${\tt SNR }_1,{\tt SNR }_2$ grow large with ${\tt SNR }_1 = \mu {\tt SNR }_2$ for some fixed ratio $0< \mu <\infty$, such that the relay gain remains finite and strictly positive.

\begin{corollary} \label{cor:asy_out}
Suppose ${\tt SNR }_1,{\tt SNR }_2$ grow large with a finite non-zero ratio and consider any independent fading distributions on $\rho_1,\rho_2$ that are strictly positive (with probability one).

The OP with AF relaying and non-ideal hardware satisfies
\begin{equation} \label{eq_pout_as_fg_af}
\lim_{{\tt SNR }_1,{\tt SNR }_2\rightarrow \infty} P_{\tt{out}}(x)  = \begin{cases} 0, & x \leq \frac{1}{\kappa_1^2 + \kappa_2^2 + \kappa_1^2\kappa_2^2},\\
1, & x > \frac{1}{\kappa_1^2 + \kappa_2^2 + \kappa_1^2\kappa_2^2}, \end{cases}
\end{equation}
while the OP with DF relaying and non-ideal hardware satisfies
\begin{equation} \label{eq_pout_as_fg_df}
\lim_{{\tt SNR }_1,{\tt SNR }_2 \rightarrow \infty} P_{\tt{out}}(x)  = \begin{cases} 0, & x \leq \frac{1}{\max(\kappa_1^2,\kappa_2^2)},\\
1, & x > \frac{1}{\max(\kappa_1^2,\kappa_2^2)}. \end{cases}
\end{equation}
\end{corollary}
\begin{IEEEproof}
Referring back to \eqref{eq_SNDR_AF_impairments_variable}, observe that we can rewrite the SNDR in terms of ${\tt SNR }_1,{\tt SNR }_2$ by extracting out the average fading power as $\rho_i = \mathbb{E}_{\rho_i}\{ \rho_i \} \tilde{\rho}_i$ (where $\tilde{\rho}_i$ represents a normalized channel gain). By taking the limit ${\tt SNR }_1,{\tt SNR }_2 \rightarrow \infty$ (with ${\tt SNR }_1 = \mu {\tt SNR }_2$), we can easily see that the end-to-end SNDR, for variable gain AF relaying, converges to
\begin{align} \label{eq:SNDR-limit}
\displaystyle\lim_{{\tt SNR }_1,{\tt SNR }_2 \rightarrow \infty} \gamma_{\nid}^{\tt{AF} \textrm{-} \tt{v}} = \frac{1}{d} =  \frac{1}{\kappa_1^2 + \kappa_2^2 + \kappa_1^2\kappa_2^2}
\end{align}
for any non-zero realization of $\tilde{\rho}_1,\tilde{\rho}_2$. Since this happens with probability one, the OP in \eqref{eq_pout_as_fg_af} is obtained in this case. The proofs for the cases of fixed gain AF relaying and DF relaying follow a similar line of reasoning.
\end{IEEEproof}

A number of conclusions can be drawn from Corollary~\ref{cor:asy_out}. First, an SNDR ceiling effect appears in the high-SNR regime,
which significantly limits the performance of both AF and DF relaying systems. This means that for $x$ smaller than the ceiling,  $P_{\tt{out}}(x) $ goes to zero with increasing SNR (at the same rate as with ideal hardware; see Section \ref{sec:results}) while the OP always equals one for $x$ larger than the ceiling. This phenomenon is fundamentally different from the ideal hardware case, in which an increasing SNR makes the end-to-end SNDR grow without bound and $P_{\tt{out}}(x) \rightarrow 0$ for any $x$. Note that this ceiling effect is independent of the fading distributions of the two hops. Similar behaviors have been observed for two-way relaying in \cite{Matthaiou2013a}, although the exact characterization is different in that configuration.

The SNDR ceiling for dual-hop relaying is
\begin{equation} \label{eq_SNDR_ceiling}
\gamma^{*} \triangleq \begin{cases} \frac{1}{\kappa_1^2 + \kappa_2^2 + \kappa_1^2\kappa_2^2} & \text{for AF protocol,} \\
\frac{1}{\max(\kappa_1^2,\kappa_2^2)} & \text{for DF protocol}, \end{cases}
\end{equation}
which is inversely proportional to the squares of $\kappa_1,\kappa_2$. This validates that transceiver hardware impairments dramatically affect the performance of relaying channels and should be taken into account when evaluating relaying systems. The ceiling is, roughly speaking, twice as large for DF relaying as for AF relaying;\footnote{This is easy to see when $\kappa_1,\kappa_2$ have the same value $\kappa>0$, which gives $\gamma^{*}=\frac{1}{\kappa^2}$ for DF relaying and
$\gamma^{*}=\frac{1}{2\kappa^2 + \kappa^4} < \frac{1}{2\kappa^2}$ for AF relaying.} this implies that the DF protocol can handle practical applications with twice as large SNDR constraints without running into a definitive outage state. Apart from this, the impact of $\kappa_1$ and $\kappa_2$ on the SNDR ceiling is similar for both relaying protocols, since $\gamma^{*}$ is a symmetric function of $\kappa_1,\kappa_2$.

We now turn our attention to the ergodic capacity in the high-SNR regime. In this case, the following result is of particular importance.

\begin{corollary}\label{asy_cap}
Suppose ${\tt SNR }_1,{\tt SNR }_2$ grow large with a finite non-zero ratio and consider any independent fading distributions on $\rho_1,\rho_2$ that are strictly positive (with probability one).

The ergodic capacity with AF relaying and non-ideal hardware satisfies
\begin{equation} \label{asy_cap_AF}
\lim_{{\tt SNR }_1,{\tt SNR }_2 \rightarrow \infty} C_{\nid}^{\tt{AF}} = \log_2\left(1+\frac{1}{\kappa_1^2 + \kappa_2^2 + \kappa_1^2\kappa_2^2} \right).
\end{equation}

The ergodic capacity with DF relaying and non-ideal hardware satisfies
\begin{equation} \label{asy_cap_DF}
\lim_{{\tt SNR }_1,{\tt SNR }_2 \rightarrow \infty} C_{\nid}^{\tt{DF}} \leq \log_2\left(1+\frac{1}{\max(\kappa_1^2,\kappa_2^2)} \right).
\end{equation}
\end{corollary}
\begin{IEEEproof}
For AF relaying, the instantaneous SNDR is upper bounded as $\gamma_{\nid}^{\tt{AF}} \leq \frac{1}{d}$ for any realizations of $\rho_1,\rho_2$. The dominated convergence theorem therefore allows us to move the limit in \eqref{asy_cap_AF} inside the expectation operator of the ergodic capacity expression in \eqref{capform}. The right-hand side of \eqref{asy_cap_AF} now follows directly from \eqref{eq:SNDR-limit}. For DF relaying, we see directly from Theorem \ref{theorem:upper-bound-capacity-DF} that $C_{\nid}^{\tt{DF}} \leq \min_i \log_2(1+1/\kappa_i^2)$, as ${\tt SNR }_i \rightarrow \infty$, which gives \eqref{asy_cap_DF}.
\end{IEEEproof}

Similar to the asymptotic OP analysis, Corollary \ref{asy_cap} demonstrates the presence of a capacity ceiling in the high-SNR regime.
This implies that transceiver hardware impairments make the ergodic capacity saturate, thereby limiting the performance of high-rate systems. Similar capacity ceilings have previously been observed for single-hop multi-antenna systems in \cite{Studer2010a,wenk2010mimo,Bjornson2013c}.
We finally point out that the approximate capacity expression in \eqref{eq:cap_fixed} becomes asymptotically exact and equal to \eqref{asy_cap_AF}, for the case of Nakagami-$m$ fading.

\subsection{Design Guidelines for Relaying Systems}
\label{subsec:design-guidelines}

Recall from Lemma \ref{lemma:single-parameter_characterization} that $\kappa_i$ is the aggregate level of impairments of the $i$th hop, for $i=1,2$. The parameter can be decomposed as
\begin{equation}
\kappa_i = \sqrt{\kappa_{i,{\tt t }}^2 + \kappa_{i,{\tt r }}^2}
\end{equation}
where $\kappa_{i,{\tt t }},\kappa_{i,{\tt r }}$ are the levels of impairments (in terms of EVM) in the transmitter and receiver hardware, respectively. The hardware cost is a decreasing function of the EVMs, because low-cost hardware has lower quality and thus higher EVMs. Hence, it is of practical interest to find the EVM combination that maximizes the performance for a fixed cost.

To provide explicit guidelines, we define the hardware cost as $\sum_{i=1}^{2} \zeta(\kappa_{i,{\tt t }}) + \zeta(\kappa_{i,{\tt r }})$, where $\zeta(\cdot)$ is a continuously decreasing, twice differentiable, and convex function. The convexity is motivated by diminishing returns; that is, high-quality hardware is more expensive to improve than low-quality hardware. The following corollary provides insights for hardware design.

\begin{corollary} \label{cor_optimal_impairments}
Suppose $\sum_{i=1}^{2} \zeta(\kappa_{i,{\tt t }}) + \zeta(\kappa_{i,{\tt r }}) = T_{\max}$ for some given cost $T_{\max} \geq 0$.
The SNDR ceilings in \eqref{eq_SNDR_ceiling} are both maximized by $ \kappa_{1,{\tt t }} = \kappa_{1,{\tt r }} = \kappa_{2,{\tt t }} = \kappa_{2,{\tt r }} = \zeta^{-1} \left( \frac{T_{\max}}{4} \right)$.
\end{corollary}
\begin{IEEEproof}
The proof goes by contradiction. Assume $\kappa_{1,{\tt t }}^*,\kappa_{1,{\tt r }}^*,\kappa_{2,{\tt t }}^*,\kappa_{2,{\tt r }}^*$ is the optimal solution and that these EVMs are not all equal. The hardware cost is a Schur-convex function (since it is convex and symmetric \cite[Proposition 2.7]{Jorswieck2007a}), thus the alternative solution $ \kappa_{1,{\tt t }} = \kappa_{1,{\tt r }} = \kappa_{2,{\tt t }} = \kappa_{2,{\tt r }} = \frac{ \sum_{i=1}^{2} \kappa_{i,{\tt t }}^*+\kappa_{i,{\tt r }}^* }{4}$ reduces the cost \cite[Theorem 2.21]{Jorswieck2007a}.
To show that the alternative solution also improves the SNDR ceilings, we first note that $\kappa_i^2 = \kappa_{i,{\tt t }}^2 + \kappa_{i,{\tt r }}^2$ is a Schur-convex function, thus it is maximized by  $\kappa_{i,{\tt t }}=\kappa_{i,{\tt r }}$ for any fixed value on $\kappa_{i,{\tt t }} + \kappa_{i,{\tt r }}$ \cite[Theorem 2.21]{Jorswieck2007a}. In addition, for any fixed value $A = \kappa_1^2 + \kappa_2^2$, $\gamma^{*}$ in \eqref{eq_SNDR_ceiling} is maximized by $\kappa_1^2= \kappa_2^2 = \frac{A}{2}$, which is easily seen from the structure of $\gamma^{*} = \frac{1}{A+A\kappa_1^2 -\kappa_1^4}$ for AF and $\gamma^{*} = \frac{1}{\max(\kappa_1^2,A-\kappa_1^2)}$ for DF. The alternative solution decreases cost and increases \eqref{eq_SNDR_ceiling}, thus the EVMs must be equal at the optimum.
\end{IEEEproof}

Corollary \ref{cor_optimal_impairments} shows that it is better to have the same level of impairments at every\footnote{There are four transceiver chains: transmitter hardware at the source, receiver and transmitter hardware at the relay, and receiver hardware at the destination.} transceiver chain, than mixing high-quality and low-quality transceiver chains. In particular, this tells us that the relay hardware should ideally be of the same quality as the source and destination hardware.

As a consequence, we provide the following design guidelines on the \emph{highest} level of impairments that can theoretically meet stipulated requirements.

\begin{corollary}  \label{cor_optimal_impairments_thresholds}
Consider a relaying system optimized according to Corollary \ref{cor_optimal_impairments}.
To support a given SNDR threshold, $x$, it is necessary to have $\kappa_i^2 \leq \sqrt{\frac{1}{x}+1}-1$ for AF relaying and $\kappa_i^2  \leq \frac{1}{x}$ for DF relaying for $i=1,2$.
\end{corollary}
\begin{IEEEproof}
Corollary \ref{cor_optimal_impairments} prescribes that $\kappa_1=\kappa_2$. Plugging this fact into \eqref{eq_SNDR_ceiling}, we obtain equations that give the expressions stated in this corollary.
\end{IEEEproof}

This corollary shows that hardware requirements are looser for DF than for AF, which is also illustrated in Section \ref{sec:results}. If the SNDR threshold is substituted as $x = 2^{2R}-1$, then we achieve the corresponding necessary conditions for achieving an ergodic capacity of $R$ bits/channel use.

Observe that the guidelines in Corollary \ref{cor_optimal_impairments_thresholds} are necessary, while the sufficiency only holds asymptotically in the high-SNR regime. Thus, practical systems should be more conservatively designed to cope with finite SNRs and different channel fading conditions.

\begin{figure}
\begin{center}
\includegraphics[width=\columnwidth]{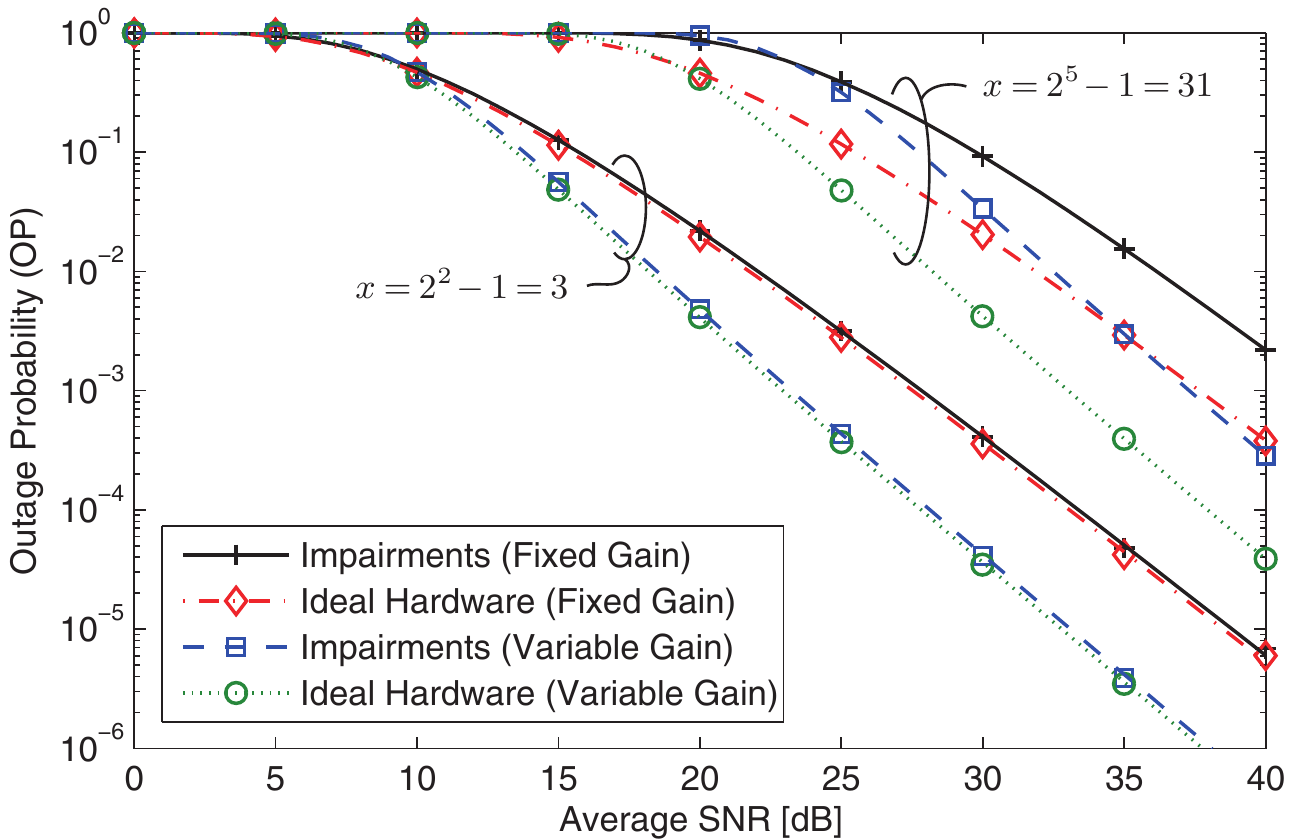}
\end{center} \vskip-3mm
\caption{Outage probability $P_{\tt{out}}(x)$ for AF relaying with ideal hardware and with hardware impairments of $\kappa_1=\kappa_2=0.1$.} \label{figure_OP_AF}
\end{figure}

\begin{figure}
\begin{center}
\includegraphics[width=\columnwidth]{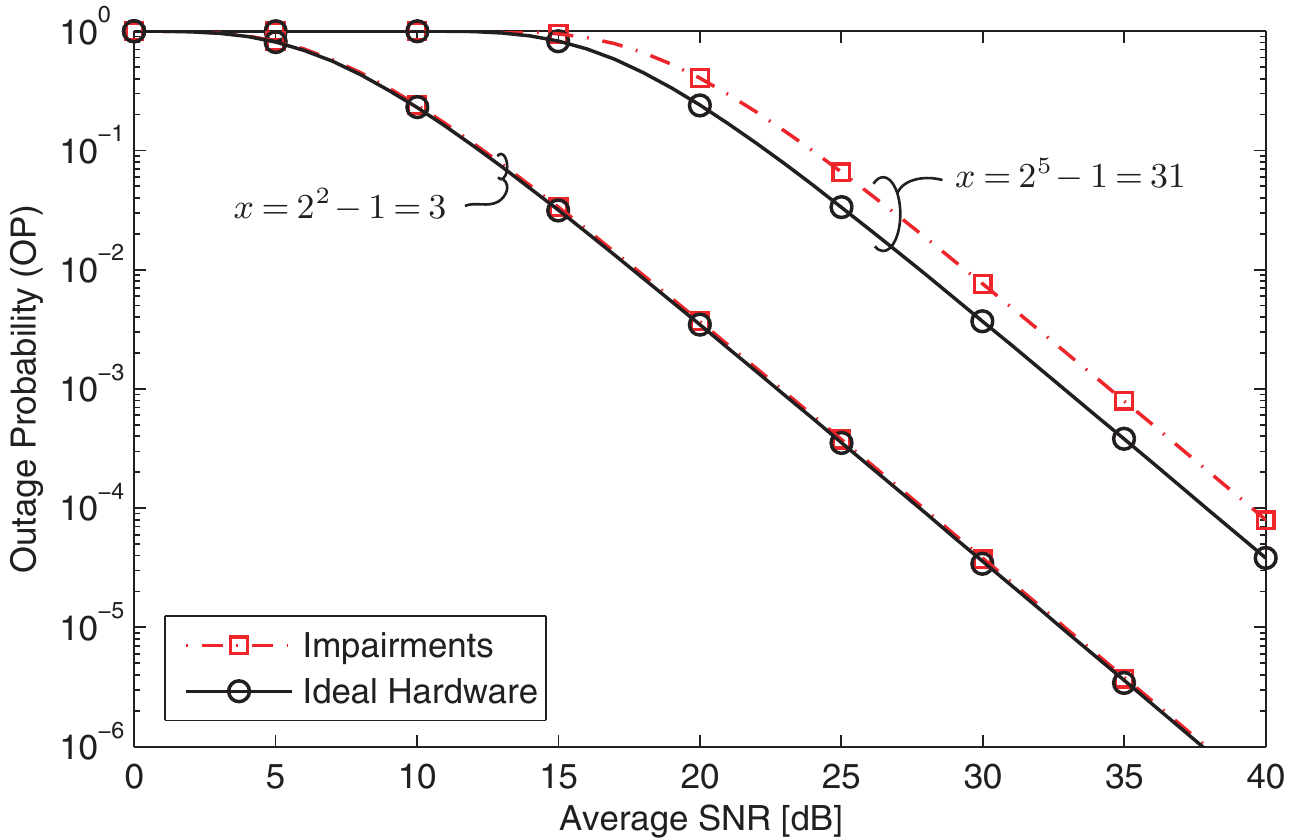}
\end{center} \vskip-3mm
\caption{Outage probability $P_{\tt{out}}(x)$ for DF relaying with ideal hardware and with hardware impairments of $\kappa_1=\kappa_2=0.1$.} \label{figure_OP_DF}
\end{figure}

\section{Numerical Illustrations}
\label{sec:results}

In this section, the theoretical results are validated by a set of Monte-Carlo simulations. Furthermore, the concepts of SNDR and capacity ceilings and the practical design guidelines of Section~\ref{section_asymptotics} are numerically illustrated.

\subsection{Different Channel Fading Conditions}

First, we consider the impact of hardware impairments on the OP, $P_{\tt{out}}(x)$, for two different thresholds: $x=2^{2}-1 =3$ and $x=2^{5}-1 =31$. Keeping in mind that the relay communication occupies two time slots, these correspond to rates of 1 and 2.5 bits/channel use, respectively. We consider a symmetric scenario with fixed levels of impairments of $\kappa_1=\kappa_2=0.1$, independent Nakagami-$m$ fading channels with $\alpha_1=\alpha_2=2$, and the same average SNR at both channels. Recall that the average SNRs are defined in \eqref{eq:SNR-definition} and note that we will not specify $\beta_1,\beta_2,P_1,P_2$ in this section since these parameters are implicitly determined by the average SNR.\footnote{Observe that while the shape parameters $\alpha_1,\alpha_2$ affect the SNR distributions of ${\tt SNR }_1$ and ${\tt SNR }_2$, respectively, any selection of the scaling parameter $\beta_i$ and the transmit power $P_i$ that gives the same value of the product $P_i \beta_i$ will give exactly the same performance and SNR distribution.} Increasing the SNR is interpreted as decreasing the propagation distance; see Remark \ref{remark:high-snr}.

Fig.~\ref{figure_OP_AF} and Fig.~\ref{figure_OP_DF} show the OP as a function of the average SNR for AF and DF relaying, respectively.
The curves in Fig.~\ref{figure_OP_AF} and Fig.~\ref{figure_OP_DF} were generated by the analytical expressions in Theorems \ref{theorem_AF_Nakagami_Rayleigh} and \ref{theorem_DF_Nakagami_Rayleigh} and show perfect agreement with the marker symbols which are the results of Monte-Carlo simulations. As shown in these figures, there is only a minor performance loss caused by transceiver hardware impairments in the low threshold case of $x=3$. However, there is a substantial performance loss when the threshold is increased to $x=31$. More precisely, AF relaying (with either variable or fixed gain) and DF relaying experience losses of around 5 dB and 2 dB in SNR, respectively, for $x=31$. The DF protocol is thus more resilient to hardware impairments, which was expected since the distortion noise of the first hop does not carry on to the second hop in this protocol. Nevertheless, the OP curves for AF and DF relaying with non-ideal hardware have the same slope as with ideal hardware; hence, hardware impairments cause merely an SNR offset that is manifested as a curve shifting to the right in Figs.~\ref{figure_OP_AF} and~\ref{figure_OP_DF}. We also note that variable gain relaying outperforms the fixed one in most scenarios of interest, which is in line with the observations in \cite{Hasna2004a}.

\begin{figure}
\begin{center}
\includegraphics[width=\columnwidth]{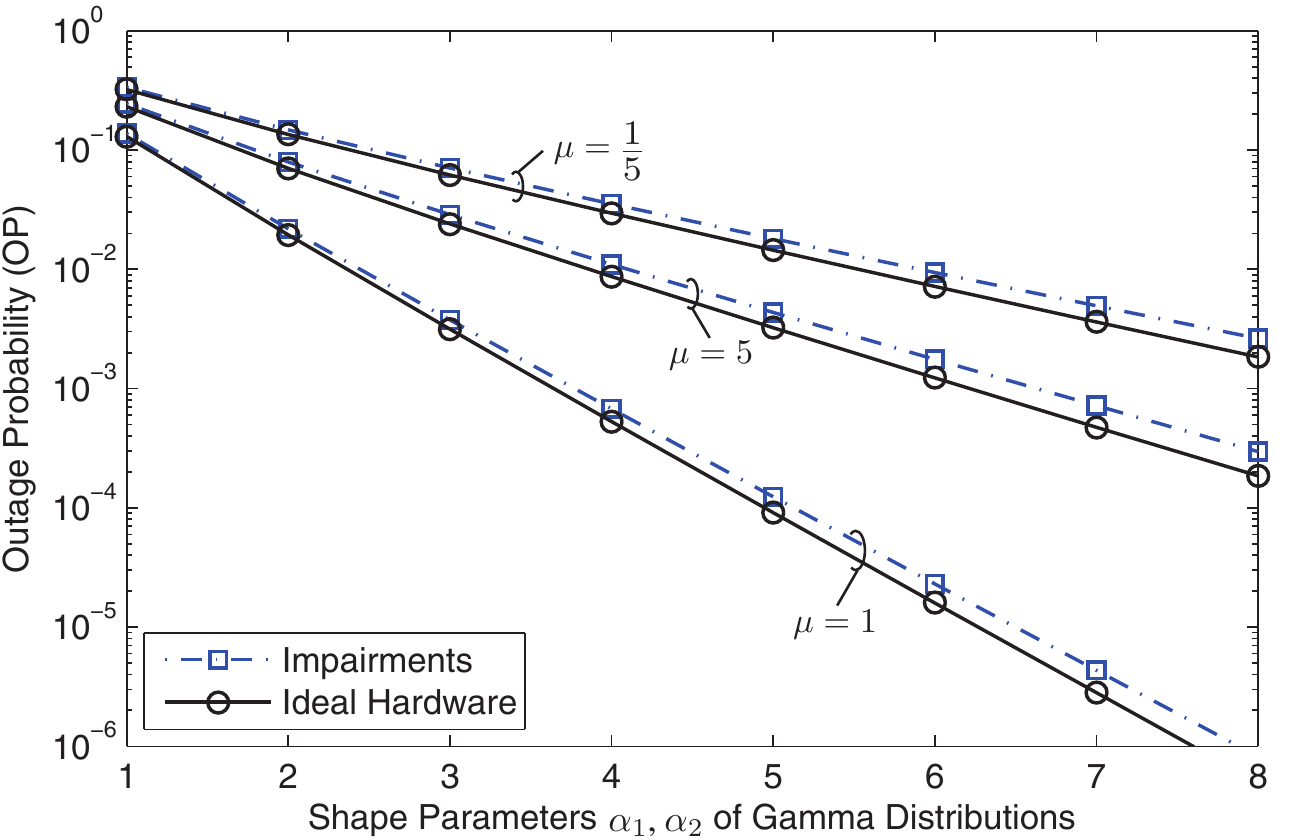}
\end{center} \vskip-3mm
\caption{Outage probability $P_{\tt{out}}(3)$ for fixed gain AF relaying with ideal hardware and with hardware impairments of $\kappa_1=\kappa_2=0.1$. Different shape parameters $\alpha_1,\alpha_2$ are considered in the fading distributions and different asymmetric SNRs: ${\tt SNR }_1= \mu{\tt SNR }_2$. The strongest channel has an SNR of 20 dB.} \label{figure_shape_parameter}
\end{figure}

Next, we illustrate the impact of the shape parameters $\alpha_1,\alpha_2$ of the Nakagami-$m$ fading distributions. We also consider different asymmetric setups where ${\tt SNR }_1= \mu {\tt SNR }_2$, for $\mu \in \{ \frac{1}{5}, \, 1, \, 5 \}$, while the largest of the SNRs is fixed as $\max({\tt SNR }_1,{\tt SNR }_2) = 20$ dB. Fig.~\ref{figure_shape_parameter} shows the OP for $x=3$ with ideal hardware and with hardware impairments characterized by $\kappa_1=\kappa_2=0.1$. We only show the result for fixed gain AF relaying for brevity. Observe that increasing the shape parameters will monotonically decrease the OP and thus improve the system performance. This is because the variance of the channel gain $\rho_i$ decreases when increasing $\alpha_i$, while we keep the average SNR fixed. Moreover, we note that it is far better to have the same SNR at both hops than asymmetries. In asymmetric cases, we note from Fig.~\ref{figure_shape_parameter} that it is better to have a strong first hop and a weak second hop than vice versa. This is explained by the amplification of noise in the AF protocol; however, this effect disappears for variable gain AF relaying and DF relaying, which is easily seen from the symmetric SNDR expressions in \eqref{eq_SNDR_AF_impairments_variable} and \eqref{eq_SNDR_DF_impairments}.

\begin{figure}
\begin{center}
\includegraphics[width=\columnwidth]{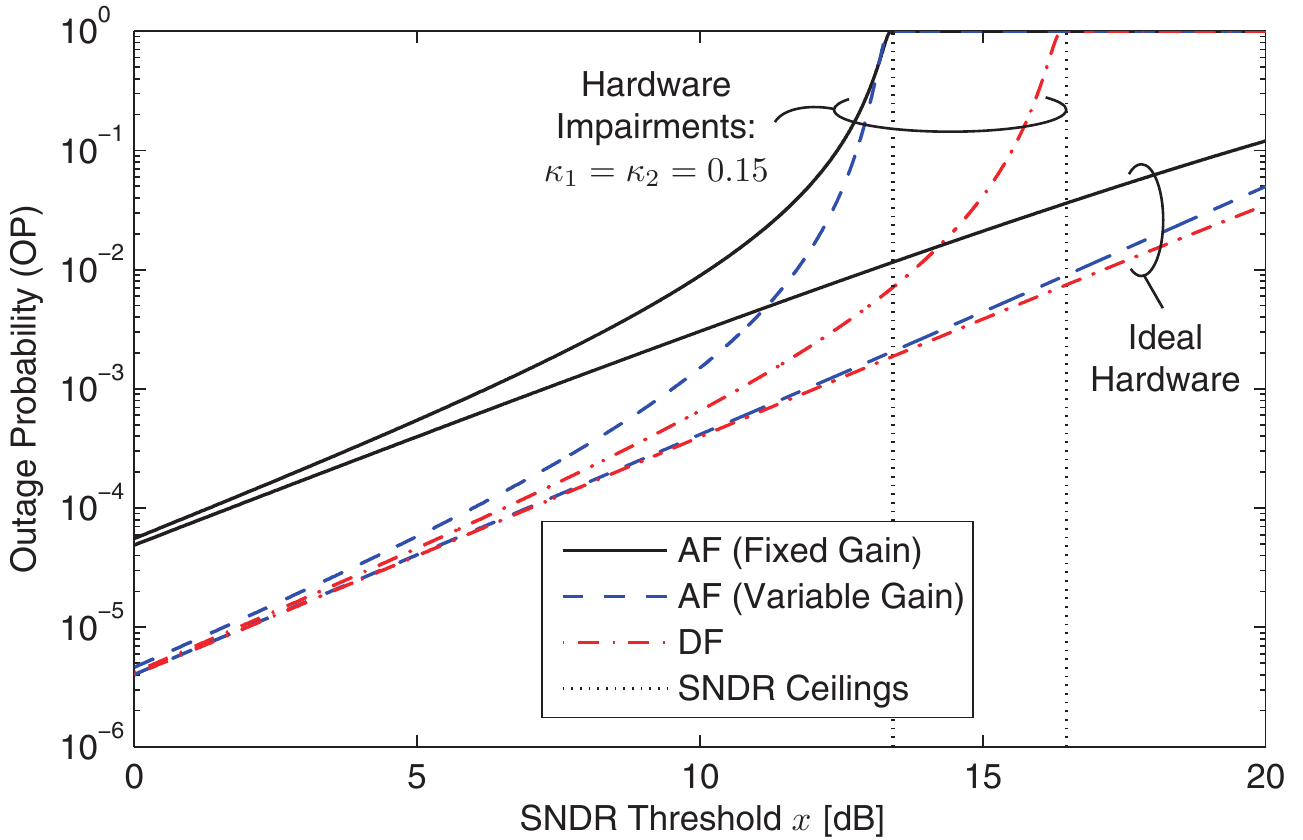}
\end{center} \vskip-3mm
\caption{Outage probability $P_{\tt{out}}(x)$ for AF and DF relaying for different thresholds $x$. As proved in Corollary \ref{cor:asy_out}, there
exist SNDR ceilings under transceiver hardware impairments.} \label{figure_SDNR_ceiling} \vskip-4mm
\end{figure}

\subsection{SNDR and Capacity Ceilings}

Next, we illustrate the existence of SNDR ceilings. To this end, we consider a fixed average SNR of 30 dB at both channels and independent Nakagami-$m$ fading channels with $\alpha_1=\alpha_2=2$. Fig.~\ref{figure_SDNR_ceiling} shows the OP, $P_{\tt{out}}(x)$, as a function of the threshold $x$ (in dB) using either ideal hardware or hardware with impairments of level $\kappa_1=\kappa_2=0.15$. For low thresholds, the OPs for AF (with fixed or variable gain) and DF are only slightly degraded by hardware impairments. The behavior is, however, very different as $x$ increases; the ideal hardware case gives a smooth convergence towards 1, while the practical case of hardware impairments experiences a quick convergence to the respective SNDR ceilings. The value of these ceilings were derived in Corollary \ref{cor:asy_out}. As noted earlier, DF relaying is more resilient to hardware impairments and its SNDR ceiling is roughly twice as large as that of AF relaying.

\begin{figure}
\begin{center}
\includegraphics[width=\columnwidth]{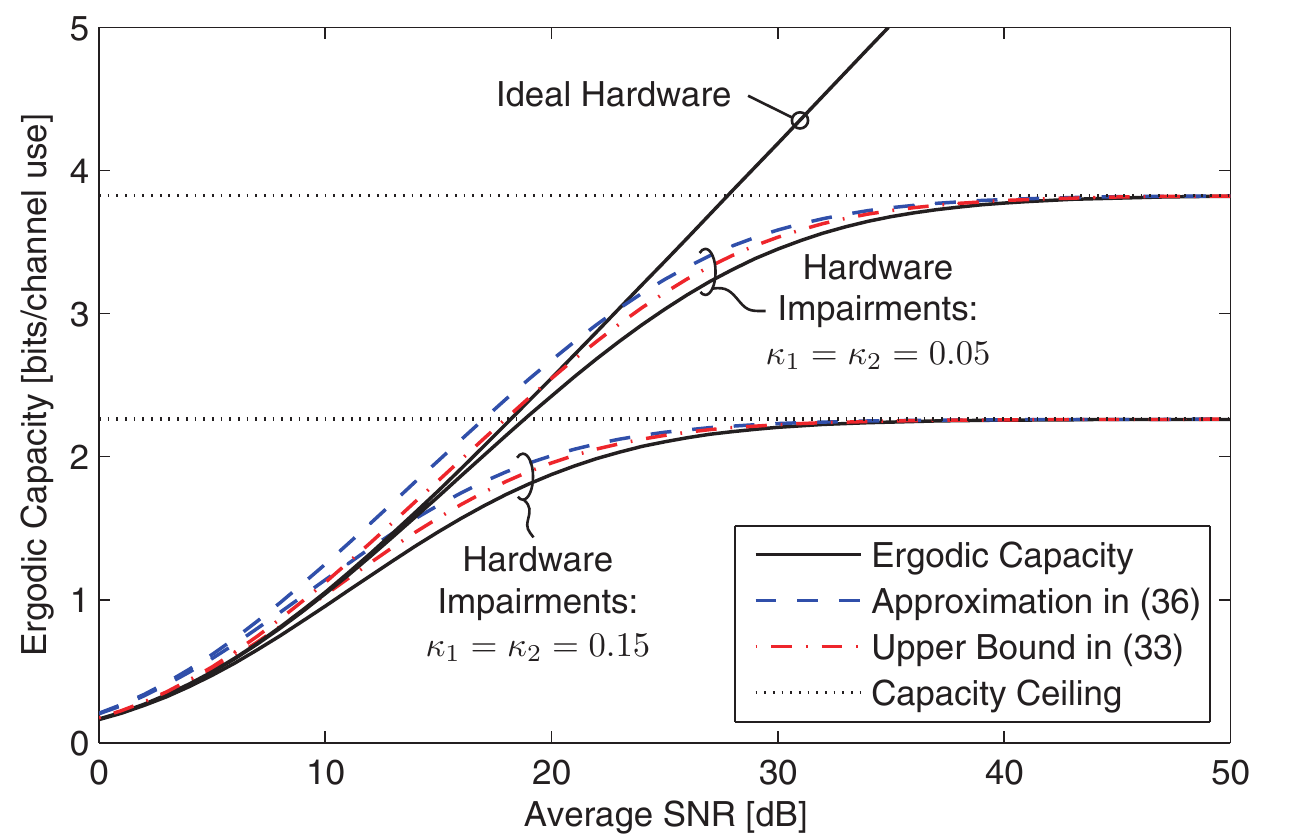}
\end{center} \vskip-3mm
\caption{Exact and approximate ergodic capacity for variable gain AF relaying. As proved in Corollary \ref{asy_cap}, there
exist capacity ceilings under transceiver hardware impairments.} \label{figure_capacity_ceiling}
\end{figure}

The similar concept of an ergodic capacity ceiling is illustrated in Fig.~\ref{figure_capacity_ceiling}, which shows the capacity of variable gain AF relaying as a function of the average SNR. Both channels are modeled as independent Nakagami-$m$ fading with $\alpha_1=\alpha_2=2$. The capacity is shown for ideal hardware and for hardware with impairments characterized by $\kappa_1=\kappa_2 \in \{ 0.05, \, 0.15 \}$. Fig.~\ref{figure_capacity_ceiling} confirms that hardware impairments have small impact at low SNRs, but are very influential at high SNRs. More precisely, the ergodic capacity saturates and approaches $\log_2 (1+ \frac{1}{\kappa_1^2 + \kappa_2^2 + \kappa_1^2\kappa_2^2})$, as proved by Corollary \ref{asy_cap}. As the capacity ceiling is determined by the level of impairments, it increases when $\kappa_1,\kappa_2$ are decreased. Fig.~\ref{figure_capacity_ceiling} also shows the upper capacity bound from Theorem \ref{theorem:upper-bound-capacity-AF} and the simplified capacity approximation from \eqref{eq:cap_fixed}. The former gives a somewhat tighter result, but both are asymptotically exact in the high-SNR regime. Although the expression \eqref{eq:cap_fixed} was derived in an approximative manner, we observe that it can indeed be considered as an upper bound on the ergodic capacity and, more importantly, is far easier to evaluate.

\begin{figure}
\begin{center}
\includegraphics[width=\columnwidth]{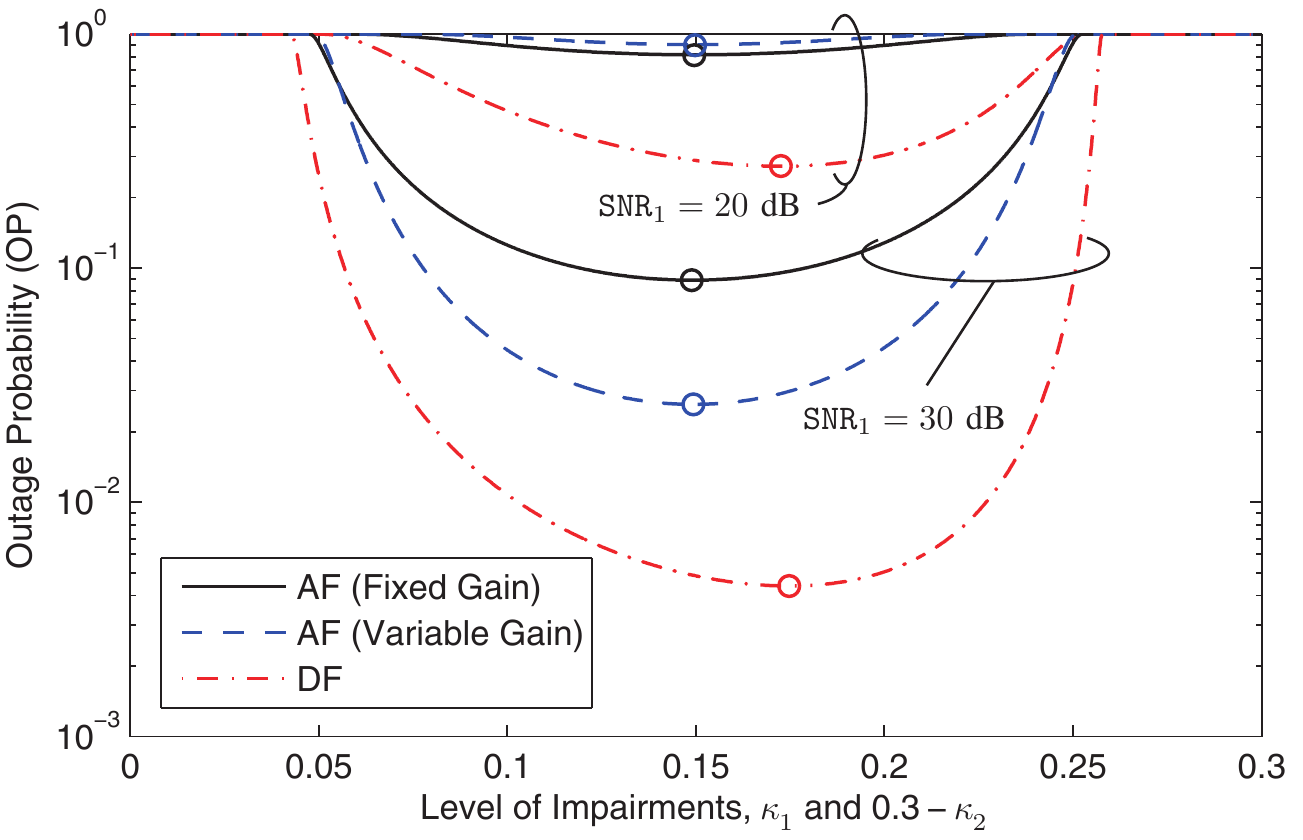}
\end{center} \vskip-3mm
\caption{Outage probability $P_{\tt{out}}(15)$  for AF and DF relaying for different levels of impairments $\kappa_1,\kappa_2$ for which $\kappa_1+\kappa_2=0.3$. The minimal value at each curve is marked with a ring.} \label{figure_designrule1}
\end{figure}

\subsection{Design Guidelines}

We conclude this section by illustrating some of the guidelines for designing practical relaying systems that were obtained in Section \ref{subsec:design-guidelines}. For simplicity, we set $\zeta(\kappa) = \kappa$ and thus limit the hardware cost by having a total EVM constraint of $\sum_{i=1}^{2} \kappa_{i,{\tt t }} + \kappa_{i,{\tt r }} =T_{\max}$. Corollary \ref{cor_optimal_impairments} proved that the SNDR ceilings are maximized by setting all $\kappa$-parameters equal to $\frac{T_{\max}}{4}$. It is intuitively clear that we should have $\kappa_{1,{\tt t }} = \kappa_{1,{\tt r }}$ and $\kappa_{2,{\tt t }} = \kappa_{2,{\tt r }}$ (see Lemma \ref{lemma:single-parameter_characterization}), but it is less obvious that the aggregate $\kappa$-parameters $\kappa_1$ and $\kappa_2$ should take the same value. To validate this property we consider an asymmetric setup where the first hop is twice as strong: ${\tt SNR }_1 = 2{\tt SNR }_2$. The channels are modeled as independent Nakagami-$m$ fading with $\alpha_1=\alpha_2=2$.

Fig.~\ref{figure_designrule1} shows the OP $P_{\tt{out}}(15)$ for two different average SNRs on the first hop: ${\tt SNR }_1 \in \{ 20, \, 30\} $ dB. The horizontal axis shows the
level of impairments of the first hop, $\kappa_1$, while the parameter of the second hop is selected to yield $\kappa_1+\kappa_2=0.3$.
Despite the asymmetric SNRs, we observe that the OP with AF relaying (with either fixed or variable gain) is minimized by setting $\kappa_1=\kappa_2 = \frac{0.3}{2}$. This shows that the design guideline in Corollary \ref{cor_optimal_impairments}, which was obtained by high-SNR analysis, can be applied successfully at finite SNRs. We also observe that the OP with DF relaying is minimized by having a slightly higher hardware quality on the weakest hop than on the strongest hop. This indicates that our general guideline should not be seen as the true optimum, but as a starting point for further adjustments. Furthermore, in the extreme cases when one of the hops is ideal ($\kappa_1 =0$ or $\kappa_2 =0$) the system is in full outage; thus, having one ideal hop does not help if the other hop has poor hardware quality.

\begin{figure}
\begin{center}
\includegraphics[width=\columnwidth]{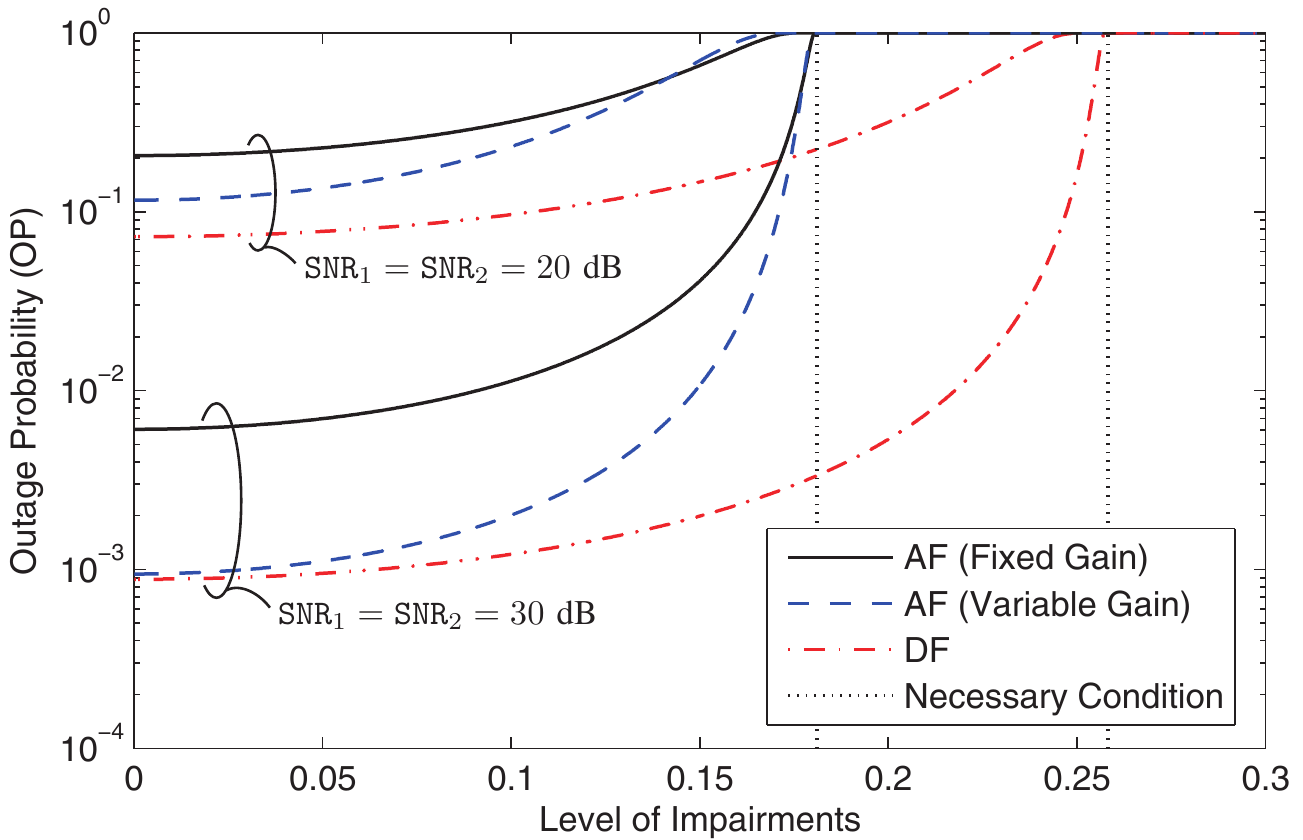}
\end{center} \vskip-3mm
\caption{Outage probability $P_{\tt{out}}(15)$ for AF and DF relaying for different symmetric levels of impairments $\kappa_1=\kappa_2$.} \label{figure_necessary_condition}
\end{figure}

Based on these insights, we now elaborate on the case with symmetric levels of impairments: $\kappa_1=\kappa_2$. Suppose our system should operate using $x=2^4-1 = 15$ (i.e., 2 bits/channel use) and we want to achieve a certain value on the outage probability $P_{\tt{out}}(15)$. Fig.~\ref{figure_necessary_condition} shows the OPs for AF and DF relaying at two different average SNRs: ${\tt SNR }_1={\tt SNR }_2 \in \{ 20, \, 30\} $ dB. Focusing on the 30 dB case and requiring that $P_{\tt{out}}(15) \leq 10^{-2}$, we can identify three possible hardware operating regimes from Fig.~\ref{figure_necessary_condition}:
\begin{enumerate}
\item Fixed gain AF relaying with $\kappa_1=\kappa_2 \leq 0.091$;
\item Variable gain AF relaying with $\kappa_1=\kappa_2 \leq 0.149$;
\item DF relaying with $\kappa_1=\kappa_2 \leq 0.218$.
\end{enumerate}
The different acceptable levels of impairments show that sophisticated protocols (AF with variable gain relaying or, preferably, DF relaying) are more robust to hardware impairments and, thus, can operate with hardware of lower quality. Fig.~\ref{figure_necessary_condition} also shows the necessary conditions of Corollary~\ref{cor_optimal_impairments_thresholds}, which act as upper bounds on the level of impairments that can possibly achieve an OP below 1. Although not sufficient, these necessary conditions provide a rough estimate of where the level of impairments must lie.

\section{Conclusions}\label{sec:conclusions}

Physical transceiver hardware introduces impairments that distort the emitted and received signals in any communication system.
While the impact of individual hardware impairments (e.g., phase noise, I/Q imbalance, and HPA non-linearities) have been well investigated
in the corresponding literature, it is the aggregate impact of all hardware impairments and the respective compensation algorithms that determine the practical system performance. Motivated by this, we considered a generalized impairment model that has been validated in prior works for single-hop communications and applied it on flat-fading dual-hop relaying, considering both AF and DF protocols. Our analytical and numerical results manifested that the performance of dual-hop relaying is notably affected by these hardware impairments, particularly when high achievable rates are required. Closed-form expressions for the exact and asymptotic OPs were derived under Nakagami-$m$ fading, along with tractable upper bounds and approximations for the ergodic capacities. These expressions effectively characterize the impact of impairments and demonstrate the existence of fundamental SNDR and capacity ceilings that cannot be crossed by increasing the signal powers or changing the fading conditions. Note that even very small hardware impairments will ultimately limit the performance. These observations also hold true for every individual subcarrier in dual-hop OFDM systems.

We finally derived some useful design guidelines for optimizing the performance of hardware-constrained relaying systems: 1) Use the same hardware quality on all transceivers; 2) Follow the necessary conditions in Corollary \ref{cor_optimal_impairments_thresholds} to find hardware qualities that can achieve the required system performance; and 3) More sophisticated relaying protocols (e.g., DF) are also more robust to hardware impairments.

\appendices

\section{Useful Lemmas}
\label{appendix:lemmas}

This appendix contains some useful lemmas. The first lemma derives the cdf of SNDR-like expressions and is used to obtain the OPs under Nakagami-$m$ fading.

\begin{lemma} \label{lemma:main-result}
Suppose $\rho_1,\rho_2$ are independent non-negative random variables with cdfs $F_{\rho_i}(\cdot)$ and pdfs $f_{\rho_i}(\cdot)$ for $i=1,2$.
Let $b_1,b_2,c,d$ be some positive scalars. The random variable
\begin{equation} \label{eq:Lambda-express}
\Lambda \triangleq \frac{\rho_1 \rho_2}{\rho_1 \rho_2 d +  \rho_1 b_1 +\rho_2  b_2 +c}
\end{equation}
has a cdf $F_{\Lambda}(x) = 0$ for $x < 0$, $F_{\Lambda}(x) = 1$ for $x \geq \frac{1}{d}$, and
\begin{equation} \label{eq:general_cdf_expression}
\begin{split}
&F_{\Lambda}  (x)  = \\ &1- \int_{0}^{\infty} \!\! \bigg( 1 \!-\! F_{\rho_1} \! \bigg( \! \frac{b_2 x}{(1 \!-\! dx)} \!+\! \frac{ \frac{b_1 b_2 x^2}{1\!-\!dx} \!+ \! cx }{z (1\!-\!dx)} \bigg) \!\! \bigg) f_{\rho_2} \! \bigg( \! z \!+\! \frac{b_1 x}{1\!-\!dx} \! \bigg) d z
\end{split}
\end{equation}
for $0 \leq x < \frac{1}{d}$. Next, let $\rho_1 \sim \mathrm{Gamma}(\alpha_1,\beta_1)$ and $\rho_2 \sim \mathrm{Gamma}(\alpha_2,\beta_2)$, where $\alpha_1,\alpha_2$ are strictly positive integers. Then, \eqref{eq:general_cdf_expression} becomes
\begin{equation} \label{eq:general_cdf_expression_gamma}
\begin{split}
F_{\Lambda} & (x)  = 1 \!-\! 2 e^{- \frac{x}{1-dx} \left(\frac{b_1}{\beta_2}+\frac{b_2}{\beta_1} \right)} \sum_{j=0}^{\alpha_1-1} \sum_{n=0}^{\alpha_2-1} \sum_{k=0}^{j} C(j,n,k) \\
&  \times \left( \frac{x}{1-dx} \right)^{\alpha_2+j} \left( b_1 b_2 + \frac{c(1-dx)}{x} \right)^{\frac{n+k+1}{2}} \\
& \times K_{n-k+1}\left( 2 \sqrt{ \frac{b_1 b_2 x^2}{\beta_1 \beta_2 (1-dx)^2} + \frac{cx}{\beta_1 \beta_2 (1-dx)} } \right)
\end{split}
\end{equation}
where $K_{\nu}(\cdot)$ denotes the $\nu$th-order modified Bessel function of the second kind and
\begin{equation}
C(j,n,k) \triangleq \frac{b_1^{\alpha_2-n-1} b_2^{j-k} \beta_1^{\frac{k-n-1-2j}{2}} \beta_2^{\frac{n-k+1-2\alpha_2}{2}} }{k! \, (j-k)! \,n! \,(\alpha_2-n-1)!}.
\end{equation}
\end{lemma}
\begin{IEEEproof}
The cdf of $\Lambda$ is defined as $F_{\Lambda}(x) = \Pr \{ \Lambda \leq x\}$. Since $\Lambda$ in \eqref{eq:Lambda-express} is a function of both $\rho_1$ and $\rho_2$, we apply the law of total probability to condition on $\rho_2$. This gives
\begin{align}
&\Pr \{ \Lambda \leq x\} = \int_{0}^{\infty} \Pr \{ \Lambda \leq x | \rho_2 \} f_{\rho_2}(\rho_2) d \rho_2 \\  \notag
 &\quad\quad \quad = 1 - \int_{0}^{\infty} \Big( 1 - \Pr \{ \Lambda \leq x | \rho_2 \} \Big) f_{\rho_2}(\rho_2) d \rho_2 \\
&= 1 \!-\! \begin{cases}
\int_{\frac{b_1 x}{1-dx}}^{\infty} \! \left( 1 - F_{\rho_1} \! \left( \frac{(b_2 \rho_2 +c) x}{\rho_2 (1-dx) - b_1 x} \right) \! \right) \! f_{\rho_2}(\rho_2) d \rho_2, &  \!\! x < \frac{1}{d}, \\
 \int_{0}^{\infty} (1 - 1) f_{\rho_2}(\rho_2) d \rho_2 = 0, & \!\!  x \geq \frac{1}{d},
\end{cases} \notag
\end{align}
where the third equality follows from evaluating the conditional probability $\Pr \{ \gamma \leq x | \rho_2 \}$ using Lemma \ref{lemma_outage_impairments}. This proves that $F_{\Lambda}(x) = 1$ for $x \geq \frac{1}{d}$. For $x < \frac{1}{d}$, we further note that
\begin{equation}
\begin{split}
&\int_{\frac{b_1 x}{1-dx}}^{\infty} \! \left( 1 - F_{\rho_1} \! \left( \frac{(b_2 \rho_2 +c) x}{\rho_2 (1\!-\!dx) - b_1 x} \right) \! \right) \! f_{\rho_2}(\rho_2) d \rho_2 \\
& \stackrel{(a)}{=} \! \int_{0}^{\infty} \!\! \bigg( 1 \!-\! F_{\rho_1} \! \bigg( \! \frac{b_2 x}{(1 \!-\! dx)} \!+\! \frac{ \frac{b_1 b_2 x^2}{1\!-\!dx} \!+ \! cx }{z (1\!-\!dx)} \bigg) \!\! \bigg) f_{\rho_2} \! \bigg( \! z \!+\! \frac{b_1 x}{1\!-\!dx} \! \bigg) d z \\
& \stackrel{(b)}{=} \! \sum_{j=0}^{\alpha_1 -1} \frac{e^{-\left( \frac{b_2 x}{\beta_1 (1 \!-\! dx)} + \frac{b_1 x}{\beta_2(1-dx)} \right) }}{j! \beta_1^{j} \beta_2^{\alpha_2} \Gamma(\alpha_2) } \! \int_{0}^{\infty} \!\! \bigg( \frac{b_2 x}{(1 \!-\! dx)} \!+\! \frac{ \frac{b_1 b_2 x^2}{1\!-\!dx} \!+ \! cx }{z (1\!-\!dx)} \bigg)^j  \\
& \quad \quad  \quad \times \bigg( \! z \!+\! \frac{b_1 x}{1\!-\!dx} \! \bigg)^{\alpha_2-1} e^{- \frac{1}{z} \left(
 \frac{ b_1 b_2 x^2}{\beta_1 (1\!-\!dx)^2} +  \frac{ cx }{\beta_1 (1\!-\!dx)} \right) - \frac{z}{\beta_2} } d z
\end{split}
\end{equation}
where $(a)$ follows from a change of variables $z = \rho_2 - \frac{b_1 x}{1-dx}$ and gives \eqref{eq:general_cdf_expression}. Furthermore, $(b)$ follows by plugging in the cdf and pdf from \eqref{eq_cdf_exp}--\eqref{eq_pdf_exp}. The remaining integral is of the form in Lemma \ref{lemma:integral-identity}. The final expression in \eqref{eq:general_cdf_expression_gamma} follows from that lemma and some algebraic simplifications.
\end{IEEEproof}

The following lemma summarizes an approach from \cite{Senaratne2010a}.

\begin{lemma} \label{lemma:integral-identity}
For any constants $c_1,c_2,c_3,c_4$ with $\Re(c_3)>0$, $\Re(c_4)>0$ and some positive integers $p_1,p_2$, we have
\begin{equation} \label{eq:integral-identity}
\begin{split}
& \int_{0}^{\infty} \big( x+c_1 \big)^{p_1} \Big( \frac{1}{x} + c_2 \Big)^{p_2} e^{-\left( \frac{c_3}{z} + z c_4 \right)} dx = 2 \sum_{n=0}^{p_1} \sum_{k=0}^{p_2} {p_1 \choose n}   \\ & \times {p_2 \choose k}  c_1^{p_1-n} c_2^{p_2-k}   \left( \frac{c_3}{c_4} \right)^{ \! \frac{n-k+1}{2}} K_{n-k+1} \left( 2 \sqrt{c_3 c_4} \right).
\end{split}
\end{equation}
Note that $\Re(\cdot)$ denotes the real part of a complex number.
\end{lemma}
\begin{IEEEproof}
The binomial formula gives the expansions
\begin{align}
\big(x+c_1 \big)^{p_1} &= \sum_{n=0}^{p_1} {p_1 \choose n} x^{n} c_1^{p_1-n} \\
\Big( \frac{1}{x} + c_2 \Big)^{p_2} &= \sum_{k=0}^{p_2} {p_2 \choose k} x^{-k} c_2^{p_2-k}
\end{align}
which transform the left-hand side of \eqref{eq:integral-identity} into
\begin{equation} \label{eq:integral-identity_proof}
\sum_{n=0}^{p_1} \sum_{k=0}^{p_2} {p_1 \choose n}  {p_2 \choose k}  c_1^{p_1-n}
c_2^{p_2-k} \int_{0}^{\infty} x^{n-k} e^{-\left( \frac{c_3}{z} + z c_4 \right)} dx.
\end{equation}
Finally, \eqref{eq:integral-identity_proof} is transformed into the right-hand side of \eqref{eq:integral-identity} by using the integral identity
\begin{equation}
\int_{0}^{\infty} \! x^{n-k} e^{-\left( \frac{c_3}{z} + z c_4\right)} dx = 2 \left( \frac{c_3}{c_4} \right)^{\! \frac{n-k+1}{2}} \!\!\! K_{n-k+1} \left( 2 \sqrt{c_3 c_4} \right)
\end{equation}
from \cite[Eq.~(3.471.9)]{Gradshteyn2007a}.
\end{IEEEproof}

\section{Proof of Theorems}

\label{appendix:proof-of-theorems}

\subsection*{Proof of Theorem \ref{theorem:upper-bound-capacity-AF}}

The end-to-end SNDRs for non-ideal hardware in \eqref{eq_SNDR_AF_impairments_fixed}--\eqref{eq_SNDR_AF_impairments_variable} are of the form
\begin{equation}
\frac{\rho_1 \rho_2}{\rho_1 \rho_2 d +  \rho_1 b_1 +\rho_2  b_2 +c} = \frac{\frac{\rho_1 \rho_2}{\rho_1 b_1 +\rho_2  b_2 +c}}{\frac{\rho_1 \rho_2}{\rho_1 b_1 +\rho_2  b_2 +c} d  +  1}.
\end{equation}
By defining $\psi \triangleq \frac{\rho_1 \rho_2}{\rho_1 b_1 +\rho_2  b_2 +c}$, it means that the ergodic capacity in \eqref{capform} is of the form $\frac{1}{2} \mathbb{E}\left\{ \log_2\left(1+  \frac{\psi}{\psi d+1} \right) \right\}$. We note that the function $\log_2\left(1+  \frac{\psi}{\psi d+1} \right)$ is concave of $\psi$ for $\psi \geq 0$, since its second derivative is
\begin{equation}
\frac{-(2 d^2 \psi+2 d (\psi+1)+1)}{(\log_e(2) (d \psi+1)^2 (d \psi+\psi+1)^2)} <0.
\end{equation}
We can therefore apply Jensen's inequality to obtain
\begin{equation}
C_{\nid}^{\tt{AF}}\! =\! \frac{1}{2} \mathbb{E} \left\{ \log_2 \Big(1+  \frac{\psi}{\psi d+1} \Big) \right\} \!\leq\!  \frac{1}{2} \log_2 \left(1\!+\!  \frac{\mathbb{E}\{\psi\}}{\mathbb{E}\{\psi\} d+1} \right).
\end{equation}
Finally, the expectation
\begin{equation}
\begin{split}
{\cal J} \triangleq \mathbb{E}\{\psi\} =  \frac{1}{b_2} \mathbb{E} \left\{ \frac{\rho_1 \rho_2}{ \rho_1 \frac{b_1}{b_2} +\rho_2  + \frac{c}{b_2}} \right\}
\end{split}
\end{equation}
equals \eqref{eq:G-expectation} by using the moment generating function derived in \cite[Theorem 3]{Senaratne2010a}.\footnote{We found a typo in \cite[Eq.~(7)]{Senaratne2010a}: the first minus sign in $M_{\Lambda}(s)$ should be a plus sign. This mistake is also seen in \cite[Fig.~3]{Senaratne2010a} where the first derivative of $M_{\Lambda}(s)$ appears to be negative, although it must be positive at $s=0$ since this represents the mean of a non-negative random variable.}

\subsection*{Proof of Theorem \ref{theorem:upper-bound-capacity-DF}}

It was shown in the proof of Theorem \ref{theorem:upper-bound-capacity-AF} above that
\begin{equation}
\mathbb{E}\left\{ \log_2\left(1+  \frac{\psi}{\psi d+1} \right) \right\} \leq  \log_2\left(1+  \frac{\mathbb{E}\{\psi\}}{\mathbb{E}\{\psi\} d+1} \right)
\end{equation}
due to Jensen's inequality and the fact that $\log_2(1+  \frac{\psi}{\psi d+1} )$ is a concave function of $\psi$ for $\psi \geq 0$. In our case, we set $\psi =  \frac{P_i \rho_i}{N_i}$, for $i=1,2$, thus $\mathbb{E}\{\psi\} = \frac{P_i \mathbb{E}_{\rho_i}\{ \rho_i \}  }{N_i} = {\tt SNR }_i$. By applying this on each expectation in \eqref{capform_DF}, we obtain \eqref{eq:cap_fixed_strict_DF}.

\section*{Acknowledgments}

The authors would like to thank Dr.~Agisilaos Papadogiannis for the indispensable discussions and collaboration that led to our joint prior works on this topic.

\bibliographystyle{IEEEbib}
\bibliography{IEEEabrv,refs}

\begin{thebibliography}{10}

\bibitem{Bjornson2013icassp}
E.~Bj\"ornson, A.~Papadogiannis, M.~Matthaiou, and M.~Debbah,
\newblock ``On the impact of transceiver impairments on {AF} relaying,''
\newblock in {\em Proc. IEEE Int. Conf. Acoustics, Speech, Signal Process.
  (ICASSP)}, May 2013.

\bibitem{Laneman2004a}
J.~N. Laneman, D.~N.~C. Tse, and G.~W. Wornell,
\newblock ``Cooperative diversity in wireless networks: Efficient protocols and
  outage behavior,''
\newblock {\em {IEEE} Trans. Inf. Theory}, vol. 50, no. 12, pp. 3062--3080,
  Dec. 2004.

\bibitem{Hasna2004a}
M.~O. Hasna and M.-S. Alouini,
\newblock ``A performance study of dual-hop transmissions with fixed gain
  relays,''
\newblock {\em {IEEE} Trans. Wireless Commun.}, vol. 3, no. 6, pp. 1963--1968,
  Nov. 2004.

\bibitem{Hasna2004b}
M.~O. Hasna and M.-S. Alouini,
\newblock ``Harmonic mean and end-to-end performance of transmission systems
  with relays,''
\newblock {\em {IEEE} Trans. Commun.}, vol. 52, no. 1, pp. 130--135, Jan. 2004.

\bibitem{Yang2009}
Y.~Yang, H.~Hu, J.~Xu, and G.~Mao,
\newblock ``Relay technologies for {WiMAX} and {LTE}-advanced mobile systems,''
\newblock {\em {IEEE} Commun. Mag.}, vol. 47, no. 10, pp. 100--105, Oct. 2009.

\bibitem{Hua2012}
Y.~Hua, D.~W. Bliss, S.~Gazor, Y.~Rong, and Y.~Sung,
\newblock ``Theories and methods for advanced wireless relays---{I}ssue {I},''
\newblock {\em {IEEE} J. Sel. Areas Commun.}, vol. 30, no. 8, pp. 1297--1303,
  Sept. 2012.

\bibitem{Costa2002a}
E.~Costa and S.~Pupolin,
\newblock ``$m$-{QAM}-{OFDM} system performance in the presence of a nonlinear
  amplifier and phase noise,''
\newblock {\em {IEEE} Trans. Commun.}, vol. 50, no. 3, pp. 462--472, Mar. 2002.

\bibitem{Schenk2008a}
T.~Schenk,
\newblock {\em {RF} Imperfections in High-Rate Wireless Systems: Impact and
  Digital Compensation},
\newblock Springer, 2008.

\bibitem{Studer2010a}
C.~Studer, M.~Wenk, and A.~Burg,
\newblock ``{MIMO} transmission with residual transmit-{RF} impairments,''
\newblock in {\em Proc.~ITG Work. Smart Ant. (WSA)}, Feb. 2010, pp. 189--196.

\bibitem{wenk2010mimo}
M.~Wenk,
\newblock {\em {MIMO-OFDM} Testbed: Challenges, Implementations, and
  Measurement Results},
\newblock Series in microelectronics. Hartung-Gorre, 2010.

\bibitem{Zetterberg2011a}
P.~Zetterberg,
\newblock ``Experimental investigation of {TDD} reciprocity-based zero-forcing
  transmit precoding,''
\newblock {\em EURASIP J. Adv. Signal Process.}, Jan. 2011.

\bibitem{schenk2007IQ}
T.~C.~W. Schenk, E.~R. Fledderus, and P.~F.~M. Smulders,
\newblock ``Performance analysis of zero-{IF} {MIMO} {OFDM} transceivers with
  {IQ} imbalance,''
\newblock {\em J. Commun.}, vol. 2, no. 7, pp. 9--19, Dec. 2007.

\bibitem{Dardari2000a}
D.~Dardari, V.~Tralli, and A.~Vaccari,
\newblock ``A theoretical characterization of nonlinear distortion effects in
  {OFDM} systems,''
\newblock {\em {IEEE} Trans. Commun.}, vol. 48, no. 10, pp. 1755--1764, Oct.
  2000.

\bibitem{Bjornson2012b}
E.~Bj{\"{o}}rnson, P.~Zetterberg, and M.~Bengtsson,
\newblock ``Optimal coordinated beamforming in the multicell downlink with
  transceiver impairments,''
\newblock in {\em Proc.~IEEE Global Commun. Conf. (GLOBECOM)}, Dec. 2012, pp.
  4775--4780.

\bibitem{Samuel2008a}
J.~Samuel, P.~Rosson, L.~Maret, C.~Dehos, and A.~Valkanas,
\newblock ``Impact of {RF} impairments in cellular wireless metropolitan area
  networks,''
\newblock in {\em Proc. IEEE Int. Symp. Spread Spectrum Techn. Appl. (ISSSTA)},
  Aug. 2008, pp. 766--769.

\bibitem{Riihonen2010a}
T.~Riihonen, S.~Werner, F.~Gregorio, R.~Wichman, and J.~H\"{a}m\"{a}l\"{a}inen,
\newblock ``{BEP} analysis of {OFDM} relay links with nonlinear power
  amplifiers,''
\newblock in {\em Proc.~IEEE Wireless Commun. Netw. Conf. (WCNC)}, Apr. 2010.

\bibitem{Qi2012a}
J.~Qi, S.~A\"{i}ssa, and M.-S. Alouini,
\newblock ``Analysis and compensation of {I/Q} imbalance in amplify-and-forward
  cooperative systems,''
\newblock in {\em Proc.~IEEE Wireless Commun. Netw. Conf. (WCNC)}, Apr. 2012,
  pp. 215--220.

\bibitem{Bjornson2013c}
E.~Bj{\"{o}}rnson, P.~Zetterberg, M.~Bengtsson, and B.~Ottersten,
\newblock ``Capacity limits and multiplexing gains of {MIMO} channels with
  transceiver impairments,''
\newblock {\em {IEEE} Commun. Lett.}, vol. 17, no. 1, pp. 91--94, Jan. 2013.

\bibitem{Bjornson2013d}
E.~Bj{\"{o}}rnson and E.~Jorswieck,
\newblock ``Optimal resource allocation in coordinated multi-cell systems,''
\newblock {\em Foundations and Trends in Communications and Information
  Theory}, vol. 9, no. 2-3, pp. 113--381, 2013.

\bibitem{Matthaiou2013a}
M.~Matthaiou, A.~Papadogiannis, E.~Bj\"{o}rnson, and M.~Debbah,
\newblock ``Two-way relaying under the presence of relay transceiver hardware
  impairments,''
\newblock {\em {IEEE} Commun. Lett.}, vol. 17, no. 6, pp. 1136--1139, Jun.
  2013.

\bibitem{Naofal1}
M.~Awadin, A.~Gomaa, and N.~Al-Dhahir,
\newblock ``{OFDM AF} relaying under {I/Q} imbalance: Performance analysis and
  baseband compensation,''
\newblock {\em {IEEE} Trans. Commun.}, vol. 61, no. 4, pp. 1304--1313, Apr.
  2013.

\bibitem{Naofal2}
O.~Ozdemir, R.~Hamila, and N.~Al-Dhahir,
\newblock ``{I/Q} imbalance in multiple beamforming {OFDM} transceivers: {SINR}
  analysis and digital baseband compensation,''
\newblock {\em {IEEE} Trans. Commun.}, vol. 61, no. 5, pp. 1914--1925, May
  2013.

\bibitem{Zhu2009a}
H.~Zhu and J.~Wang,
\newblock ``Chunk-based resource allocation in {OFDMA} systems---{P}art {I}:
  {C}hunk allocation,''
\newblock {\em {IEEE} Trans. Commun.}, vol. 57, no. 9, pp. 2734--2744, Sept.
  2009.

\bibitem{Priyanto2007a}
B.~E. Priyanto, T.~B. Sorensen, O.~K. Jensen, T.~Larsem, T.~Kolding, and
  P.~Mogensen,
\newblock ``Assessing and modelling the effect of {RF} impairments on {UTRA}
  {LTE} uplink performance,''
\newblock in {\em Proc.~IEEE Vehic. Techn. Conf. (VTC)}, Sept. 2007, pp.
  1213--1217.

\bibitem{Donoughue2012a}
N.~O'Donoughue and J.~M.~F. Moura,
\newblock ``On the product of independent complex {Gaussians},''
\newblock {\em {IEEE} Trans. Signal Process.}, vol. 60, no. 3, pp. 1050--1063,
  Mar. 2012.

\bibitem{EVM2005}
``8 hints for making and interpreting {EVM} measurements,''
\newblock Tech. {R}ep., Agilent Technologies, 2005.

\bibitem{Holma2011a}
H.~Holma and A.~Toskala,
\newblock {\em {LTE} for {UMTS}: {Evolution} to {LTE}-Advanced},
\newblock Wiley, 2nd edition, 2011.

\bibitem{Sheikh1993}
A.~U. Sheikh, M.~Handforth, and M.~Abdi,
\newblock ``Indoor mobile radio channel at 946 {MHz}: Measurements and
  modeling,''
\newblock in {\em Proc.~IEEE Veh. Techn. Conf. (VTC)}, May 1993, pp. 73--76.

\bibitem{Studer2011a}
C.~Studer, M.~Wenk, and A.~Burg,
\newblock ``System-level implications of residual transmit-{RF} impairments in
  {MIMO} systems,''
\newblock in {\em Proc.~Europ. Conf. Ant. Propag. (EuCAP)}, Apr. 2011, pp.
  2686--2689.

\bibitem{Emamian2002a}
V.~Emamian, P.~Anghel, and M.~Kaveh,
\newblock ``Multi-user spatial diversity in a shadow-fading environment,''
\newblock in {\em Proc.~IEEE Veh. Techn. Conf. (VTC)}, Sept. 2002, pp.
  573--576.

\bibitem{Karagiannidis2006a}
G.~K. Karagiannidis, T.~A. Tsiftsis, and R.~K. Mallik,
\newblock ``Bounds for multihop relayed communications in {Nakagami}-$m$
  fading,''
\newblock {\em {IEEE} Trans. Commun.}, vol. 54, no. 1, pp. 130--135, Jan. 2006.

\bibitem{Senaratne2010a}
D.~Senaratne and C.~Tellambura,
\newblock ``Unified exact performance analysis of two-hop amplify-and-forward
  relaying in {Nakagami} fading,''
\newblock {\em {IEEE} Trans. Veh. Technol.}, vol. 59, no. 3, pp. 1529--1534,
  Mar. 2010.

\bibitem{Hasna2002}
M.~O. Hasna and M.-S. Alouini,
\newblock ``Performance analysis of two-hop relayed transmissions over
  {R}ayleigh fading channels,''
\newblock in {\em Proc.~IEEE Veh. Techn. Conf. (VTC)}, Sept. 2002, pp.
  1992--1996.

\bibitem{Zhao2005}
Y.~Zhao, R.~Adve, and T.~J. Lim,
\newblock ``Outage probability at arbitrary {SNR} with cooperative diversity,''
\newblock {\em {IEEE} Commun. Lett.}, vol. 9, no. 8, pp. 700--702, Aug. 2005.

\bibitem{Farhadi2008a}
G.~Farhadi and N.~C. Beaulieu,
\newblock ``On the ergodic capacity of wireless relaying systems over
  {Rayleigh} fading channels,''
\newblock {\em {IEEE} Trans. Wireless Commun.}, vol. 7, no. 11, pp. 4462--4467,
  Nov. 2008.

\bibitem{Waqar2011}
O.~Waqar, M.~Ghogho, and D.~McLernon,
\newblock ``Tight bounds for ergodic capacity of dual-hop fixed-gain relay
  networks under {R}ayleigh fading,''
\newblock {\em {IEEE} Commun. Lett.}, vol. 15, no. 4, pp. 413--415, Apr. 2011.

\bibitem{Zhong2011}
C.~Zhong, M.~Matthaiou, G.~K. Karagiannidis, and T.~Ratnarajah,
\newblock ``Generic ergodic capacity bounds for fixed-gain {AF} dual-hop
  relaying systems,''
\newblock {\em {IEEE} Trans. Veh. Technol.}, vol. 60, no. 8, pp. 3814--3824,
  Oct. 2011.

\bibitem{Host2005a}
A.~H{\o}st-Madsen and J.~Zhang,
\newblock ``Capacity bounds and power allocation for wireless relay channels,''
\newblock {\em {IEEE} Trans. Inf. Theory}, vol. 51, no. 6, pp. 2020--2040, Jun.
  2005.

\bibitem{Gradshteyn2007a}
I.~S. Gradshteyn and I.~M. Ryzhik,
\newblock {\em Table of Integrals, Series, and Products},
\newblock Academic Press, 7th edition, 2007.

\bibitem{Weiss2010a}
A.~Weiss,
\newblock ``Optimization using symbolic derivatives,''
\newblock Tech. {R}ep., MATLAB Digest, 2010.

\bibitem{Jorswieck2007a}
E.~Jorswieck and H.~Boche,
\newblock ``Majorization and matrix-monotone functions in wireless
  communications,''
\newblock {\em Foundations and Trends in Communications and Information
  Theory}, vol. 3, no. 6, pp. 553--701, 2007.

\end{thebibliography}

\vspace{-5mm}

\begin{IEEEbiography}[{\includegraphics[width=1in,height=1.25in,clip,keepaspectratio]{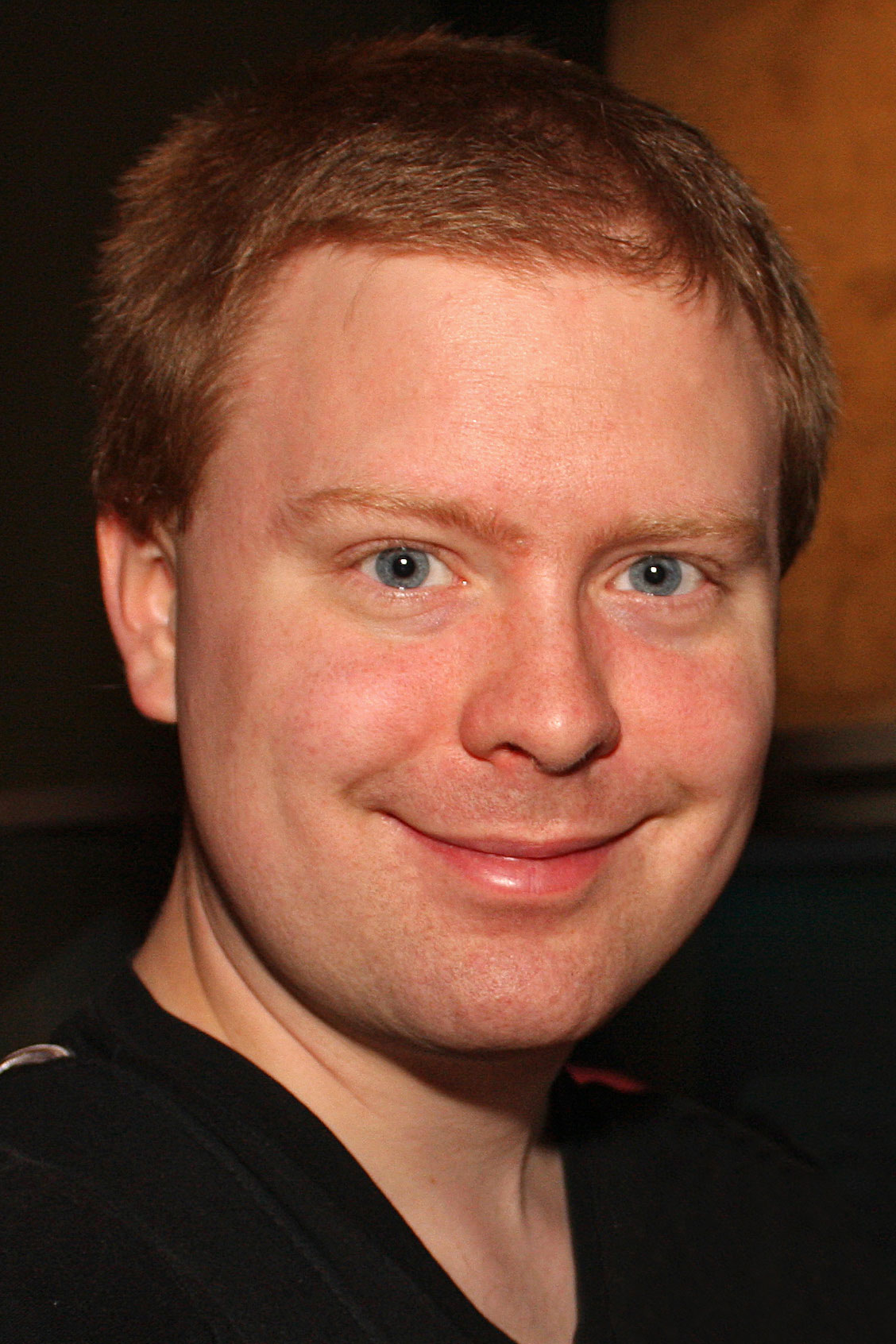}}]{Emil Bj\"{o}rnson}
(S'07--M'12) was born in Malm\"{o}, Sweden, in 1983. He received the M.S. degree in Engineering Mathematics from Lund University, Lund, Sweden, in 2007. He received the Ph.D. degree in Telecommunications from the Signal Processing Lab at KTH Royal Institute of Technology, Stockholm, Sweden, in 2011. He is the first author of the monograph ``Optimal Resource Allocation in Coordinated Multi-Cell Systems'' published in Foundations and Trends in Communications and Information Theory, January 2013.

Dr. Bj\"{o}rnson was one of the first recipients of the International Postdoc Grant from the Swedish Research Council. This grant is currently funding a joint postdoctoral research fellowship at the Alcatel-Lucent Chair on Flexible Radio, Sup\'{e}lec, Paris, France, and the Signal Processing Lab at KTH Royal Institute of Technology, Stockholm, Sweden. His research interests include multi-antenna cellular communications, resource allocation, random matrix theory, estimation theory, stochastic signal processing, and mathematical optimization.

For his work on optimization of multi-cell MIMO communications, he received a Best Paper Award at the 2009 International Conference on Wireless Communications \& Signal Processing (WCSP) and a Best Student Paper Award at the 2011 IEEE International Workshop on Computational Advances in Multi-Sensor Adaptive Processing (CAMSAP).
\end{IEEEbiography}

\begin{biography}[{\includegraphics[width=1in,height=1.25in,clip,keepaspectratio]{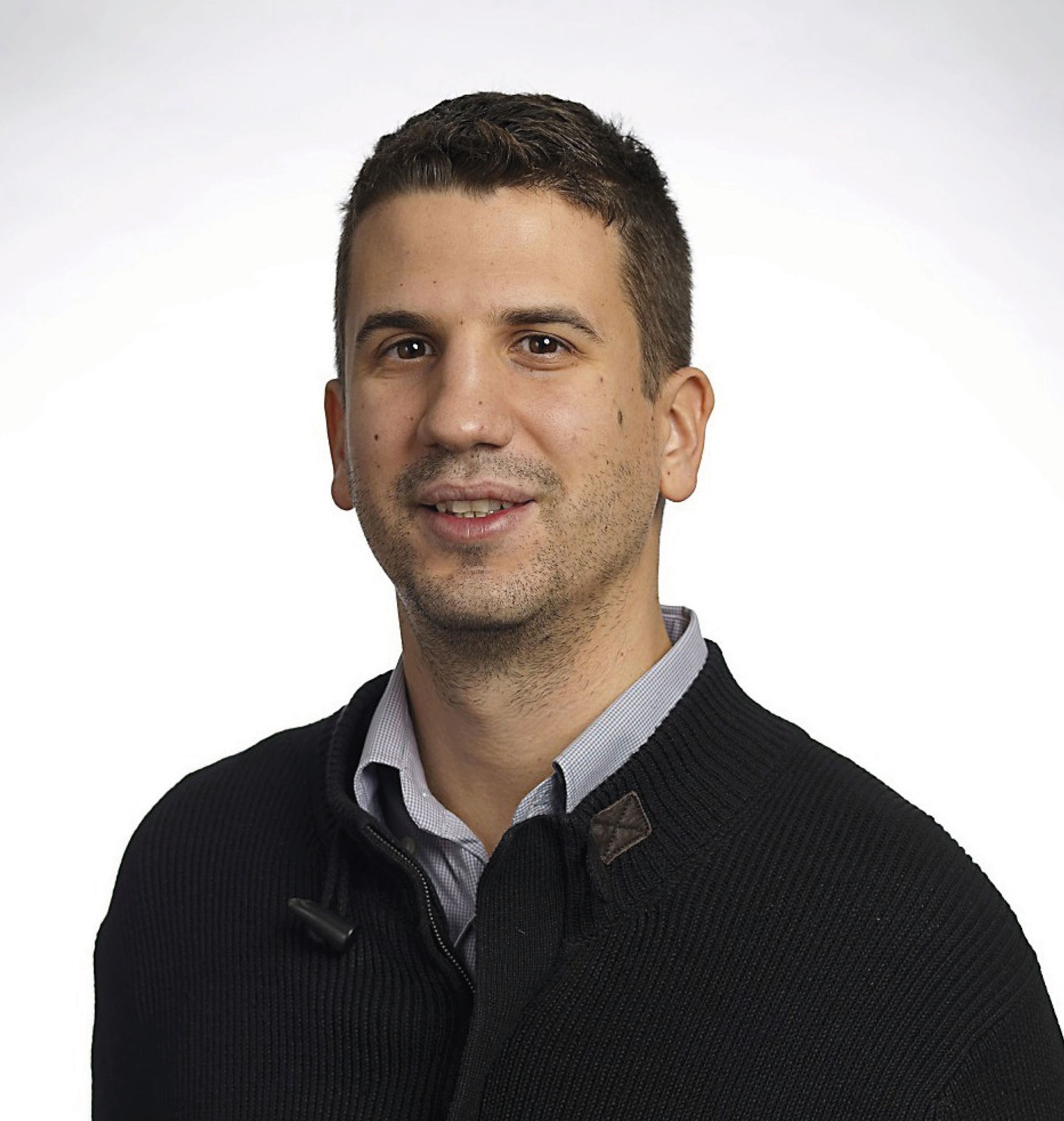}}]
{Michail Matthaiou}(S'05--M'08--SM'13) was born in Thessaloniki, Greece in 1981. He obtained the Diploma degree (5 years) in Electrical and Computer Engineering from the Aristotle University of Thessaloniki, Greece in 2004. He then received the M.Sc. (with distinction) in Communication Systems and Signal Processing from the University of Bristol, U.K. and Ph.D. degrees from the University of Edinburgh, U.K. in 2005 and 2008, respectively. From September 2008 through May 2010, he was with the Institute for Circuit Theory and Signal Processing, Munich University of Technology (TUM), Germany working as a Postdoctoral Research Associate. In June 2010, he joined Chalmers University of Technology, Sweden as an Assistant Professor and in 2011 he was awarded the Docent title. His research interests span signal processing for wireless communications, random matrix theory and multivariate statistics for MIMO systems, and performance analysis of fading channels.

Dr. Matthaiou is the recipient of the 2011 IEEE ComSoc Young Researcher Award for the Europe, Middle East and Africa Region and a co-recipient of the 2006 IEEE Communications Chapter Project Prize for the best M.Sc. dissertation in the area of communications. He was an Exemplary Reviewer for \textsc{IEEE Communications Letters} for 2010. He has been a member of  Technical Program Committees for several IEEE conferences such as ICC, GLOBECOM, etc. He currently serves as an Associate Editor for the \textsc{IEEE Transactions on Communications}, \textsc{IEEE Communications Letters} and was the Lead Guest Editor of the special issue on ``Large-scale multiple antenna wireless systems'' of the \textsc{IEEE Journal on Selected Areas in Communications}. He is an associate member of the IEEE Signal Processing Society SPCOM and SAM technical committees.
\end{biography}

\newpage

\begin{biography}[{\includegraphics[width=1in,height=1.25in,clip,keepaspectratio]{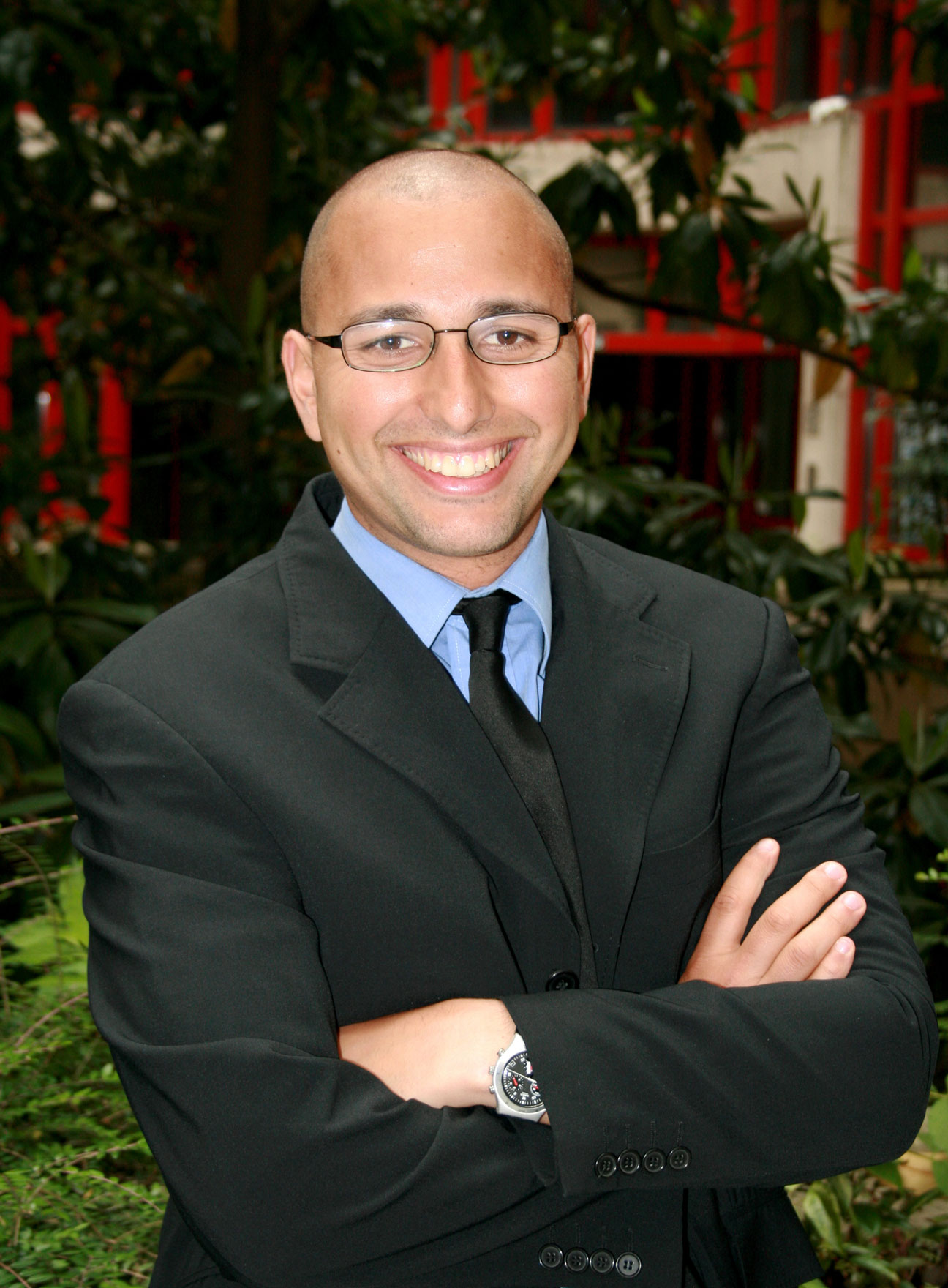}}]
{M\'{e}rouane Debbah}(SM'08) entered the Ecole Normale Sup\'{e}rieure de Cachan (France) in 1996 where he received his M.Sc. and Ph.D. degrees respectively. He worked for Motorola Labs (Saclay, France) from 1999-2002 and the Vienna Research Center for Telecommunications (Vienna, Austria) until 2003. He then joined the Mobile Communications department of the Institut Eurecom (Sophia Antipolis, France) as an Assistant Professor until 2007. He is now a Full Professor at Supel\'{e}c (Gif-sur-Yvette, France), holder of the Alcatel-Lucent Chair on Flexible Radio and a recipient of the ERC starting grant MORE (Advanced Mathematical Tools for Complex Network Engineering).

His research interests are in information theory, signal processing and wireless communications. He is a senior area editor for IEEE Transactions on Signal Processing and an Associate Editor in Chief of the journal Random Matrix: Theory and Applications. M\'{e}rouane Debbah is the recipient of the "Mario Boella" award in 2005, the 2007 General Symposium IEEE GLOBECOM best paper award, the Wi-Opt 2009 best paper award, the 2010 Newcom++ best paper award, the WUN CogCom Best Paper 2012 and 2013 Award as well as the Valuetools 2007, Valuetools 2008, Valuetools 2012 and CrownCom2009 best student paper awards. He is a WWRF fellow and an elected member of the academic senate of Paris-Saclay. In 2011, he received the IEEE Glavieux Prize Award.
\end{biography}

\end{document}